\documentclass[
]{ceurart}

\usepackage{listings}
\usepackage{numprint}
\npthousandsep{,}
\usepackage{tikz}
\usetikzlibrary{positioning, shapes, arrows.meta, bending}
\usepackage{mathtools}

\newcommand{\SPXXXVIII}{
  $-4$ & \texttt{-{}-{}-{}-{}-{}-{}-{}-{}-{}-{}-{}-{}-{}-{}-{}-{}-{}-{}-{}-{}-{}-{}-{}-{}-{}-{}-{}-{}-{}-{}-{}-{}} & \texttt{-{}-{}-{}-{}-{}-{}-{}-{}-{}-{}-{}-{}-{}-{}-{}-{}-{}-{}-{}-{}-{}-{}-{}-{}-{}-{}-{}-{}-{}-{}-{}-{}} &  \\
  $-3$ & \texttt{-{}-{}-{}-{}-{}-{}-{}-{}-{}-{}-{}-{}-{}-{}-{}-{}-{}-{}-{}-{}-{}-{}-{}-{}-{}-{}-{}-{}-{}-{}-{}-{}} & \texttt{-{}-{}-{}-{}-{}-{}-{}-{}-{}-{}-{}-{}-{}-{}-{}-{}-{}-{}-{}-{}-{}-{}-{}-{}-{}-{}-{}-{}-{}-{}-{}-{}} &  \\
  $-2$ & \texttt{-{}-{}-{}-{}-{}-{}-{}-{}-{}-{}-{}-{}-{}-{}-{}-{}-{}-{}-{}-{}-{}-{}-{}-{}-{}-{}-{}-{}-{}-{}-{}-{}} & \texttt{-{}-{}-{}-{}-{}-{}-{}-{}-{}-{}-{}-{}-{}-{}-{}-{}-{}-{}-{}-{}-{}-{}-{}-{}-{}-{}-{}-{}-{}-{}-{}-{}} &  \\
  $-1$ & \texttt{-{}-{}-{}-{}-{}-{}-{}-{}-{}-{}-{}-{}-{}-{}-{}-{}-{}-{}-{}-{}-{}-{}-{}-{}-{}-{}-{}-{}-{}-{}-{}-{}} & \texttt{-{}-{}-{}-{}-{}-{}-{}-{}-{}-{}-{}-{}-{}-{}-{}-{}-{}-{}-{}-{}-{}-{}-{}-{}-{}-{}-{}-{}-{}-{}-{}-{}} &  \\
  $0$ & \texttt{-{}-{}-{}-{}-{}-{}-{}-{}-{}-{}-{}-{}-{}-{}-{}-{}-{}-{}-{}-{}-{}-{}-{}-{}-{}-{}-{}-{}-{}-{}-{}-{}} & \texttt{-{}-{}-{}-{}-{}-{}-{}-{}-{}-{}-{}-{}-{}-{}-{}-{}-{}-{}-{}-{}-{}-{}-{}-{}-{}-{}-{}-{}-{}-{}-{}-{}} & \texttt{-{}-{}-{}-{}-{}-{}-{}-{}-{}-{}-{}-{}-{}-{}-{}-{}-{}-{}-{}-{}-{}-{}-{}-{}-{}-{}-{}-{}-{}-{}-{}-{}} \\
  $1$ & \texttt{-{}-{}-{}-{}-{}-{}-{}-{}-{}-{}-{}-{}-{}-{}-{}-{}-{}-{}-{}-{}-{}-{}-{}-{}-{}-{}-{}-{}-{}-{}-{}-{}} & \texttt{-{}-{}-{}-{}-{}-{}-{}-{}-{}-{}-{}-{}-{}-{}-{}-{}-{}-{}-{}-{}-{}-{}-{}-{}-{}-{}-{}-{}-{}-{}-{}-{}} & \texttt{-{}-{}-{}-{}-{}-{}-{}-{}-{}-{}-{}-{}-{}-{}-{}-{}-{}-{}-{}-{}-{}-{}-{}-{}-{}-{}-{}-{}-{}-{}-{}-{}} \\
  $2$ & \texttt{-{}-{}-{}-{}-{}-{}-{}-{}-{}-{}-{}-{}-{}-{}-{}-{}-{}-{}-{}-{}-{}-{}-{}-{}-{}-{}-{}-{}-{}-{}-{}-{}} & \texttt{-{}-{}-{}-{}-{}-{}-{}-{}-{}-{}-{}-{}-{}-{}-{}-{}-{}-{}-{}-{}-{}-{}-{}-{}-{}-{}-{}-{}-{}-{}-{}-{}} & \texttt{-{}-{}-{}-{}-{}-{}-{}-{}-{}-{}-{}-{}-{}-{}-{}-{}-{}-{}-{}-{}-{}-{}-{}-{}-{}-{}-{}-{}-{}-{}-{}-{}} \\
  $3$ & \texttt{-{}-{}-{}-{}-{}-{}-{}-{}-{}-{}-{}-{}-{}-{}-{}-{}-{}-{}-{}-{}-{}-{}-{}-{}-{}-{}-{}-{}-{}-{}-{}-{}} & \texttt{-{}-{}-{}-{}-{}-{}-{}-{}-{}-{}-{}-{}-{}-{}-{}-{}-{}-{}-{}-{}-{}-{}-{}-{}-{}-{}-{}-{}-{}-{}-{}-{}} & \texttt{-{}-{}-{}-{}-{}-{}-{}-{}-{}-{}-{}-{}-{}-{}-{}-{}-{}-{}-{}-{}-{}-{}-{}-{}-{}-{}-{}-{}-{}-{}-{}-{}} \\
  $4$ & \texttt{-{}-{}-{}-{}-{}-{}-{}-{}-{}-{}-{}-{}-{}-{}-{}-{}-{}-{}-{}-{}-{}-{}-{}-{}-{}-{}-{}-{}-{}-{}-{}-{}} & \texttt{-{}-{}-{}-{}-{}-{}-{}-{}-{}-{}-{}-{}-{}-{}-{}-{}-{}-{}-{}-{}-{}-{}-{}-{}-{}-{}-{}-{}-{}-{}-{}-{}} & \texttt{-{}-{}-{}-{}-{}-{}-{}-{}-{}-{}-{}-{}-{}-{}-{}-{}-{}-{}-{}-{}-{}-{}-{}-{}-{}-{}-{}-{}-{}-{}-{}-{}} \\
  $5$ & \texttt{-{}-{}-{}-{}-{}-{}-{}-{}-{}-{}-{}-{}-{}-{}-{}-{}-{}-{}-{}-{}-{}-{}-{}-{}-{}-{}-{}-{}-{}-{}-{}-{}} & \texttt{-{}-{}-{}-{}-{}-{}-{}-{}-{}-{}-{}-{}-{}-{}-{}-{}-{}-{}-{}-{}-{}-{}-{}-{}-{}-{}-{}-{}-{}-{}-{}-{}} & \texttt{-{}-{}-{}-{}-{}-{}-{}-{}-{}-{}-{}-{}-{}-{}-{}-{}-{}-{}-{}-{}-{}-{}-{}-{}-{}-{}-{}-{}-{}-{}-{}-{}} \\
  $6$ & \texttt{-{}-{}-{}-{}-{}-{}-{}-{}-{}-{}-{}-{}-{}-{}-{}-{}-{}-{}-{}-{}-{}-{}-{}-{}-{}-{}-{}-{}-{}-{}-{}-{}} & \texttt{-{}-{}-{}-{}-{}-{}-{}-{}-{}-{}-{}-{}-{}-{}-{}-{}-{}-{}-{}-{}-{}-{}-{}-{}-{}-{}-{}-{}-{}-{}-{}-{}} & \texttt{-{}-{}-{}-{}-{}-{}-{}-{}-{}-{}-{}-{}-{}-{}-{}-{}-{}-{}-{}-{}-{}-{}-{}-{}-{}-{}-{}-{}-{}-{}-{}-{}} \\
  $7$ & \texttt{?{}?{}?{}?{}?{}?{}?{}?{}?{}?{}?{}?{}?{}?{}?{}?{}?{}?{}?{}?{}?{}?{}?{}?{}?{}?{}?{}?{}?{}?{}?{}?{}} & \texttt{?{}?{}?{}?{}?{}?{}?{}?{}?{}?{}?{}?{}?{}?{}?{}?{}?{}?{}?{}?{}?{}?{}?{}?{}?{}?{}?{}?{}?{}?{}?{}?{}} & \texttt{?{}?{}?{}?{}?{}?{}?{}?{}?{}?{}?{}?{}?{}?{}?{}?{}?{}?{}?{}?{}?{}?{}?{}?{}?{}?{}?{}?{}?{}?{}?{}?{}} \\
  $8$ & \texttt{?{}?{}?{}?{}?{}?{}?{}?{}?{}?{}?{}?{}?{}?{}?{}?{}?{}?{}?{}?{}?{}?{}?{}?{}?{}?{}?{}?{}?{}?{}?{}?{}} & \texttt{?{}?{}?{}?{}?{}?{}?{}?{}?{}?{}?{}?{}?{}?{}?{}?{}?{}?{}?{}?{}?{}?{}?{}?{}?{}?{}?{}?{}?{}?{}?{}?{}} & \texttt{?{}?{}?{}?{}?{}?{}?{}?{}?{}?{}?{}?{}?{}?{}?{}?{}?{}?{}?{}?{}?{}?{}?{}?{}?{}?{}?{}?{}?{}?{}?{}?{}} \\
  $9$ & \texttt{?{}?{}?{}?{}?{}?{}?{}?{}?{}?{}?{}?{}?{}?{}?{}?{}?{}?{}?{}?{}?{}?{}?{}?{}?{}?{}?{}?{}?{}?{}?{}?{}} & \texttt{?{}?{}?{}?{}?{}?{}?{}?{}?{}?{}?{}?{}?{}?{}?{}?{}?{}?{}?{}?{}?{}?{}?{}?{}?{}?{}?{}?{}?{}?{}?{}?{}} & \texttt{-{}-{}-{}-{}-{}-{}-{}-{}-{}-{}-{}-{}-{}-{}-{}-{}-{}-{}-{}-{}-{}-{}-{}-{}-{}-{}-{}-{}-{}-{}-{}-{}} \\
  $10$ & \texttt{-{}-{}-{}-{}-{}-{}-{}-{}-{}-{}-{}-{}-{}-{}-{}-{}-{}-{}-{}-{}-{}-{}-{}-{}-{}-{}-{}-{}-{}-{}-{}-{}} & \texttt{?{}?{}?{}?{}?{}?{}?{}?{}?{}?{}?{}?{}?{}?{}?{}?{}?{}?{}?{}?{}?{}?{}?{}?{}?{}?{}?{}?{}?{}?{}?{}?{}} & \texttt{?{}?{}?{}?{}?{}?{}?{}?{}?{}?{}?{}?{}?{}?{}?{}?{}?{}?{}?{}?{}?{}?{}?{}?{}?{}?{}?{}?{}?{}?{}?{}?{}} \\
  $11$ & \texttt{-{}-{}-{}-{}-{}-{}-{}-{}-{}-{}-{}-{}-{}-{}-{}-{}-{}-{}-{}-{}-{}-{}-{}-{}-{}-{}-{}-{}-{}-{}-{}-{}} & \texttt{?{}?{}?{}?{}?{}?{}?{}?{}?{}?{}?{}?{}?{}?{}?{}?{}?{}?{}?{}?{}?{}?{}?{}?{}?{}?{}?{}?{}?{}?{}?{}?{}} & \texttt{-{}-{}-{}-{}-{}-{}-{}-{}-{}-{}-{}-{}-{}-{}-{}-{}-{}-{}-{}-{}-{}-{}-{}-{}-{}-{}-{}-{}-{}-{}-{}-{}} \\
  $12$ & \texttt{-{}-{}-{}-{}-{}-{}-{}-{}-{}-{}-{}-{}-{}-{}-{}-{}-{}-{}-{}-{}-{}-{}-{}-{}-{}-{}-{}-{}-{}-{}-{}-{}} & \texttt{?{}?{}?{}?{}?{}?{}?{}?{}?{}?{}?{}?{}?{}?{}?{}?{}?{}?{}?{}?{}?{}?{}?{}?{}?{}?{}?{}?{}?{}?{}?{}?{}} & \texttt{-{}-{}-{}-{}-{}-{}-{}-{}-{}-{}-{}-{}-{}-{}-{}-{}-{}-{}-{}-{}-{}-{}-{}-{}-{}-{}-{}-{}-{}-{}-{}-{}} \\
  $13$ & \texttt{-{}-{}-{}-{}-{}-{}-{}-{}-{}-{}-{}-{}-{}-{}-{}-{}-{}-{}-{}-{}-{}-{}-{}-{}-{}-{}-{}-{}-{}-{}-{}-{}} & \texttt{?{}?{}?{}?{}?{}?{}?{}?{}?{}?{}?{}?{}?{}?{}?{}?{}?{}?{}?{}?{}?{}?{}?{}?{}?{}?{}?{}?{}?{}?{}?{}?{}} & \texttt{-{}-{}-{}-{}-{}-{}-{}-{}-{}-{}-{}-{}-{}-{}-{}-{}-{}-{}-{}-{}-{}-{}-{}-{}-{}-{}-{}-{}-{}-{}-{}-{}} \\
  $14$ & \texttt{-{}-{}-{}-{}-{}-{}-{}-{}-{}-{}-{}-{}-{}-{}-{}-{}-{}-{}-{}-{}-{}-{}-{}-{}-{}-{}-{}-{}-{}-{}-{}-{}} & \texttt{?{}?{}?{}?{}?{}?{}?{}?{}?{}?{}?{}?{}?{}?{}?{}?{}?{}?{}?{}?{}?{}?{}?{}?{}?{}?{}?{}?{}?{}?{}?{}?{}} & \texttt{-{}-{}-{}-{}-{}-{}-{}-{}-{}-{}-{}-{}-{}-{}-{}-{}-{}-{}-{}-{}-{}-{}-{}-{}-{}-{}-{}-{}-{}-{}-{}-{}} \\
  $15$ & \texttt{-{}-{}-{}-{}x{}-{}-{}-{}-{}-{}-{}-{}-{}-{}-{}x{}-{}-{}-{}-{}-{}-{}-{}-{}-{}-{}-{}-{}-{}-{}-{}-{}} & \texttt{?{}?{}?{}?{}?{}?{}?{}?{}?{}?{}?{}?{}?{}?{}?{}?{}?{}?{}?{}?{}?{}?{}?{}?{}?{}?{}?{}?{}?{}?{}?{}?{}} & \texttt{-{}-{}-{}-{}-{}-{}-{}-{}-{}-{}-{}-{}-{}-{}-{}-{}-{}-{}-{}-{}-{}-{}-{}-{}-{}-{}-{}-{}-{}x{}-{}-{}} \\
  $16$ & \texttt{-{}-{}-{}-{}-{}-{}-{}-{}-{}-{}-{}-{}-{}-{}-{}-{}-{}-{}-{}-{}-{}-{}-{}-{}-{}-{}-{}-{}-{}x{}-{}-{}} & \texttt{?{}?{}?{}?{}?{}?{}?{}?{}?{}?{}?{}?{}?{}?{}?{}?{}?{}?{}?{}?{}?{}?{}?{}?{}?{}?{}?{}?{}?{}?{}?{}?{}} & \texttt{-{}-{}-{}-{}-{}-{}-{}-{}-{}-{}-{}-{}-{}-{}-{}-{}-{}-{}-{}-{}-{}-{}-{}-{}-{}-{}-{}-{}-{}-{}-{}-{}} \\
  $17$ & \texttt{-{}-{}-{}-{}-{}-{}-{}-{}-{}-{}-{}-{}-{}-{}-{}-{}-{}-{}-{}-{}-{}-{}-{}-{}-{}-{}-{}-{}-{}-{}-{}-{}} & \texttt{?{}?{}?{}?{}?{}?{}?{}?{}?{}?{}?{}?{}?{}?{}?{}?{}?{}?{}?{}?{}?{}?{}?{}?{}?{}?{}?{}?{}?{}?{}?{}?{}} & \texttt{-{}-{}-{}-{}-{}-{}-{}-{}-{}-{}-{}-{}-{}-{}-{}-{}-{}-{}-{}-{}-{}-{}-{}-{}-{}-{}-{}-{}-{}-{}-{}-{}} \\
  $18$ & \texttt{-{}-{}-{}-{}-{}-{}-{}-{}-{}-{}-{}-{}-{}-{}-{}-{}-{}-{}-{}-{}-{}-{}-{}-{}-{}-{}-{}-{}-{}-{}-{}-{}} & \texttt{-{}-{}-{}-{}-{}-{}-{}-{}-{}-{}-{}-{}-{}-{}-{}-{}-{}-{}-{}-{}-{}-{}-{}-{}-{}-{}-{}-{}-{}-{}-{}-{}} & \texttt{-{}-{}-{}-{}-{}-{}-{}-{}-{}-{}-{}-{}-{}-{}-{}-{}-{}-{}-{}-{}-{}-{}-{}-{}-{}-{}-{}-{}-{}-{}-{}-{}} \\
  $19$ & \texttt{-{}-{}-{}-{}-{}-{}-{}-{}-{}-{}-{}-{}-{}-{}-{}-{}-{}-{}-{}-{}-{}-{}-{}-{}-{}-{}-{}-{}-{}-{}-{}-{}} & \texttt{?{}?{}?{}?{}?{}?{}?{}?{}?{}?{}?{}?{}?{}?{}?{}?{}?{}?{}?{}?{}?{}?{}?{}?{}?{}?{}?{}?{}?{}?{}?{}?{}} & \texttt{-{}-{}-{}-{}-{}-{}-{}-{}-{}-{}-{}-{}-{}-{}-{}-{}-{}-{}-{}-{}-{}-{}-{}-{}-{}-{}-{}-{}-{}-{}-{}-{}} \\
  $20$ & \texttt{-{}-{}-{}-{}-{}-{}-{}-{}-{}-{}-{}-{}-{}-{}-{}-{}-{}-{}-{}-{}-{}-{}-{}-{}-{}-{}-{}-{}-{}-{}-{}-{}} & \texttt{?{}?{}?{}?{}?{}?{}?{}?{}?{}?{}?{}?{}?{}?{}?{}?{}?{}?{}?{}?{}?{}?{}?{}?{}?{}?{}?{}?{}?{}?{}?{}?{}} & \texttt{-{}-{}-{}-{}-{}-{}-{}-{}-{}-{}-{}-{}-{}-{}-{}-{}-{}-{}-{}-{}-{}-{}-{}-{}-{}-{}-{}-{}-{}-{}-{}-{}} \\
  $21$ & \texttt{-{}-{}-{}-{}-{}-{}-{}-{}-{}-{}-{}-{}-{}-{}-{}-{}-{}-{}-{}-{}-{}-{}-{}-{}-{}-{}-{}-{}-{}-{}-{}-{}} & \texttt{-{}-{}-{}-{}-{}-{}-{}-{}-{}-{}-{}-{}-{}-{}-{}-{}-{}-{}-{}-{}-{}-{}-{}-{}-{}-{}-{}-{}-{}-{}-{}-{}} & \texttt{-{}-{}-{}-{}-{}-{}-{}-{}-{}-{}-{}-{}-{}-{}-{}-{}-{}-{}-{}-{}-{}-{}-{}-{}-{}-{}-{}-{}-{}-{}-{}-{}} \\
  $22$ & \texttt{-{}-{}-{}-{}-{}-{}-{}-{}-{}-{}-{}-{}-{}-{}-{}-{}-{}-{}-{}-{}-{}-{}-{}-{}-{}-{}-{}-{}-{}-{}-{}-{}} & \texttt{-{}-{}-{}-{}-{}-{}-{}-{}-{}-{}-{}-{}-{}-{}-{}-{}-{}-{}-{}-{}-{}-{}-{}-{}-{}-{}-{}-{}-{}-{}-{}-{}} & \texttt{-{}-{}-{}-{}-{}-{}-{}-{}-{}-{}-{}-{}-{}-{}-{}-{}-{}-{}-{}-{}-{}-{}-{}-{}-{}-{}-{}-{}-{}-{}-{}-{}} \\
  $23$ & \texttt{-{}-{}-{}-{}-{}-{}-{}-{}-{}-{}-{}-{}-{}-{}-{}-{}-{}-{}-{}-{}-{}-{}-{}-{}-{}-{}-{}-{}-{}-{}-{}-{}} & \texttt{-{}-{}-{}-{}-{}-{}-{}-{}-{}-{}-{}-{}-{}-{}-{}-{}-{}-{}-{}-{}-{}-{}-{}-{}-{}-{}-{}-{}-{}-{}-{}-{}} & \texttt{-{}-{}-{}-{}x{}-{}-{}-{}-{}-{}-{}-{}-{}-{}x{}x{}-{}-{}-{}-{}-{}-{}-{}-{}-{}-{}-{}-{}-{}-{}-{}-{}} \\
  $24$ & \texttt{-{}-{}-{}-{}-{}-{}-{}-{}-{}-{}-{}-{}-{}-{}-{}-{}-{}-{}-{}-{}-{}-{}-{}-{}-{}-{}-{}-{}-{}-{}-{}-{}} & \texttt{-{}-{}-{}-{}-{}-{}-{}-{}-{}-{}-{}-{}-{}-{}-{}-{}-{}-{}-{}-{}-{}-{}-{}-{}-{}-{}-{}-{}-{}-{}-{}-{}} & \texttt{-{}-{}-{}-{}-{}-{}-{}-{}-{}-{}-{}-{}-{}-{}-{}-{}-{}-{}-{}-{}-{}-{}-{}-{}-{}-{}-{}-{}-{}x{}-{}-{}} \\
  $25$ & \texttt{-{}-{}-{}-{}-{}-{}-{}-{}-{}-{}-{}-{}-{}-{}-{}-{}-{}-{}-{}-{}-{}-{}-{}-{}-{}-{}-{}-{}-{}-{}-{}-{}} & \texttt{-{}-{}-{}-{}-{}-{}-{}-{}-{}-{}-{}-{}-{}-{}-{}-{}-{}-{}-{}-{}-{}-{}-{}-{}-{}-{}-{}-{}-{}-{}-{}-{}} & \texttt{-{}-{}-{}-{}-{}-{}-{}-{}-{}-{}-{}-{}-{}-{}-{}-{}-{}-{}-{}-{}-{}-{}-{}-{}-{}-{}-{}-{}-{}-{}-{}-{}} \\
  $26$ & \texttt{-{}-{}-{}-{}-{}-{}-{}-{}-{}-{}-{}-{}-{}-{}-{}-{}-{}-{}-{}-{}-{}-{}-{}-{}-{}-{}-{}-{}-{}-{}-{}-{}} & \texttt{-{}-{}-{}-{}-{}-{}-{}-{}-{}-{}-{}-{}-{}-{}-{}-{}-{}-{}-{}-{}-{}-{}-{}-{}-{}-{}-{}-{}-{}-{}-{}-{}} & \texttt{-{}-{}-{}-{}-{}-{}-{}-{}-{}-{}-{}-{}-{}-{}-{}-{}-{}-{}-{}-{}-{}-{}-{}-{}-{}-{}-{}-{}-{}-{}-{}-{}} \\
  $27$ & \texttt{-{}-{}-{}-{}-{}-{}-{}-{}-{}-{}-{}-{}-{}-{}-{}-{}-{}-{}-{}-{}-{}-{}-{}-{}-{}-{}-{}-{}-{}-{}-{}-{}} & \texttt{-{}-{}-{}-{}-{}-{}-{}-{}-{}-{}-{}-{}-{}-{}-{}-{}-{}-{}-{}-{}-{}-{}-{}-{}-{}-{}-{}-{}-{}-{}-{}-{}} & \texttt{-{}-{}-{}-{}-{}-{}-{}-{}-{}-{}-{}-{}-{}-{}-{}-{}-{}-{}-{}-{}-{}-{}-{}-{}-{}-{}-{}-{}-{}-{}-{}-{}} \\
  $28$ & \texttt{-{}-{}-{}-{}-{}-{}-{}-{}-{}-{}-{}-{}-{}-{}-{}-{}-{}-{}-{}-{}-{}-{}-{}-{}-{}-{}-{}-{}-{}-{}-{}-{}} & \texttt{-{}-{}-{}-{}-{}-{}-{}-{}-{}-{}-{}-{}-{}-{}-{}-{}-{}-{}-{}-{}-{}-{}-{}-{}-{}-{}-{}-{}-{}-{}-{}-{}} & \texttt{-{}-{}-{}-{}-{}-{}-{}-{}-{}-{}-{}-{}-{}-{}-{}-{}-{}-{}-{}-{}-{}-{}-{}-{}-{}-{}-{}-{}-{}-{}-{}-{}} \\
  $29$ & \texttt{-{}-{}-{}-{}-{}-{}-{}-{}-{}-{}-{}-{}-{}-{}-{}-{}-{}-{}-{}-{}-{}-{}-{}-{}-{}-{}-{}-{}-{}-{}-{}-{}} & \texttt{-{}-{}-{}-{}-{}-{}-{}-{}-{}-{}-{}-{}-{}-{}-{}-{}-{}-{}-{}-{}-{}-{}-{}-{}-{}-{}-{}-{}-{}-{}-{}-{}} & \texttt{-{}-{}-{}-{}-{}-{}-{}-{}-{}-{}-{}-{}-{}-{}-{}-{}-{}-{}-{}-{}-{}-{}-{}-{}-{}-{}-{}-{}-{}-{}-{}-{}} \\
  $30$ & \texttt{-{}-{}-{}-{}-{}-{}-{}-{}-{}-{}-{}-{}-{}-{}-{}-{}-{}-{}-{}-{}-{}-{}-{}-{}-{}-{}-{}-{}-{}-{}-{}-{}} & \texttt{-{}-{}-{}-{}-{}-{}-{}-{}-{}-{}-{}-{}-{}-{}-{}-{}-{}-{}-{}-{}-{}-{}-{}-{}-{}-{}-{}-{}-{}-{}-{}-{}} & \texttt{-{}-{}-{}-{}-{}-{}-{}-{}-{}-{}-{}-{}-{}-{}-{}-{}-{}-{}-{}-{}-{}-{}-{}-{}-{}-{}-{}-{}-{}-{}-{}-{}} \\
  $31$ & \texttt{-{}-{}-{}-{}-{}-{}-{}-{}-{}-{}-{}-{}-{}-{}-{}-{}-{}-{}-{}-{}-{}-{}-{}-{}-{}-{}-{}-{}-{}-{}-{}-{}} & \texttt{-{}-{}-{}-{}-{}-{}-{}-{}-{}-{}-{}-{}-{}-{}-{}-{}-{}-{}-{}-{}-{}-{}-{}-{}-{}-{}-{}-{}-{}-{}-{}-{}} & \texttt{-{}-{}-{}-{}-{}-{}-{}-{}-{}-{}-{}-{}-{}-{}-{}-{}-{}-{}-{}-{}-{}-{}-{}-{}-{}-{}-{}-{}-{}-{}-{}-{}} \\
  $32$ & \texttt{-{}-{}-{}-{}-{}-{}-{}-{}-{}-{}-{}-{}-{}-{}-{}-{}-{}-{}-{}-{}-{}-{}-{}-{}-{}-{}-{}-{}-{}-{}-{}-{}} & \texttt{-{}-{}-{}-{}-{}-{}-{}-{}-{}-{}-{}-{}-{}-{}-{}-{}-{}-{}-{}-{}-{}-{}-{}-{}-{}-{}-{}-{}-{}-{}-{}-{}} & \texttt{-{}-{}-{}-{}-{}-{}-{}-{}-{}-{}-{}-{}-{}-{}-{}-{}-{}-{}-{}-{}-{}-{}-{}-{}-{}-{}-{}-{}-{}-{}-{}-{}} \\
  $33$ & \texttt{-{}-{}-{}-{}-{}-{}-{}-{}-{}-{}-{}-{}-{}-{}-{}-{}-{}-{}-{}-{}-{}-{}-{}-{}-{}-{}-{}-{}-{}-{}-{}-{}} & \texttt{-{}-{}-{}-{}-{}-{}-{}-{}-{}-{}-{}-{}-{}-{}-{}-{}-{}-{}-{}-{}-{}-{}-{}-{}-{}-{}-{}-{}-{}-{}-{}-{}} & \texttt{-{}-{}-{}-{}-{}-{}-{}-{}-{}-{}-{}-{}-{}-{}-{}-{}-{}-{}-{}-{}-{}-{}-{}-{}-{}-{}-{}-{}-{}-{}-{}-{}} \\
  $34$ & \texttt{-{}-{}-{}-{}-{}-{}-{}-{}-{}-{}-{}-{}-{}-{}-{}-{}-{}-{}-{}-{}-{}-{}-{}-{}-{}-{}-{}-{}-{}-{}-{}-{}} & \texttt{-{}-{}-{}-{}-{}-{}-{}-{}-{}-{}-{}-{}-{}-{}-{}-{}-{}-{}-{}-{}-{}-{}-{}-{}-{}-{}-{}-{}-{}-{}-{}-{}} & \texttt{-{}-{}-{}-{}-{}-{}-{}-{}-{}-{}-{}-{}-{}-{}-{}-{}-{}-{}-{}-{}-{}-{}-{}-{}-{}-{}-{}-{}-{}-{}-{}-{}} \\
  $35$ & \texttt{-{}-{}-{}-{}-{}-{}-{}-{}-{}-{}-{}-{}-{}-{}-{}-{}-{}-{}-{}-{}-{}-{}-{}-{}-{}-{}-{}-{}-{}-{}-{}-{}} & \texttt{-{}-{}-{}-{}-{}-{}-{}-{}-{}-{}-{}-{}-{}-{}-{}-{}-{}-{}-{}-{}-{}-{}-{}-{}-{}-{}-{}-{}-{}-{}-{}-{}} & \texttt{-{}-{}-{}-{}-{}-{}-{}-{}-{}-{}-{}-{}-{}-{}-{}-{}-{}-{}-{}-{}-{}-{}-{}-{}-{}-{}-{}-{}-{}-{}-{}-{}} \\
  $36$ & \texttt{-{}-{}-{}-{}-{}-{}-{}-{}-{}-{}-{}-{}-{}-{}-{}-{}-{}-{}-{}-{}-{}-{}-{}-{}-{}-{}-{}-{}-{}-{}-{}-{}} & \texttt{-{}-{}-{}-{}-{}-{}-{}-{}-{}-{}-{}-{}-{}-{}-{}-{}-{}-{}-{}-{}-{}-{}-{}-{}-{}-{}-{}-{}-{}-{}-{}-{}} & \texttt{-{}-{}-{}-{}-{}-{}-{}-{}-{}-{}-{}-{}-{}-{}-{}-{}-{}-{}-{}-{}-{}-{}-{}-{}-{}-{}-{}-{}-{}-{}-{}-{}} \\
  $37$ & \texttt{-{}-{}-{}-{}-{}-{}-{}-{}-{}-{}-{}-{}-{}-{}-{}-{}-{}-{}-{}-{}-{}-{}-{}-{}-{}-{}-{}-{}-{}-{}-{}-{}} & \texttt{-{}-{}-{}-{}-{}-{}-{}-{}-{}-{}-{}-{}-{}-{}-{}-{}-{}-{}-{}-{}-{}-{}-{}-{}-{}-{}-{}-{}-{}-{}-{}-{}} & \texttt{-{}-{}-{}-{}-{}-{}-{}-{}-{}-{}-{}-{}-{}-{}-{}-{}-{}-{}-{}-{}-{}-{}-{}-{}-{}-{}-{}-{}-{}-{}-{}-{}} \\
}

\newcommand{\SPXXVIII}{
  $-4$ & \texttt{-{}-{}-{}-{}-{}-{}-{}-{}-{}-{}-{}-{}-{}-{}-{}-{}-{}-{}-{}-{}-{}-{}-{}-{}-{}-{}-{}-{}-{}-{}-{}-{}} & \texttt{-{}-{}-{}-{}-{}-{}-{}-{}-{}-{}-{}-{}-{}-{}-{}-{}-{}-{}-{}-{}-{}-{}-{}-{}-{}-{}-{}-{}-{}-{}-{}-{}} &  \\
  $-3$ & \texttt{-{}-{}-{}-{}-{}-{}-{}-{}-{}-{}-{}-{}-{}-{}-{}-{}-{}-{}-{}-{}-{}-{}-{}-{}-{}-{}-{}-{}-{}-{}-{}-{}} & \texttt{-{}-{}-{}-{}-{}-{}-{}-{}-{}-{}-{}-{}-{}-{}-{}-{}-{}-{}-{}-{}-{}-{}-{}-{}-{}-{}-{}-{}-{}-{}-{}-{}} &  \\
  $-2$ & \texttt{-{}-{}-{}-{}-{}-{}-{}-{}-{}-{}-{}-{}-{}-{}-{}-{}-{}-{}-{}-{}-{}-{}-{}-{}-{}-{}-{}-{}-{}-{}-{}-{}} & \texttt{-{}-{}-{}-{}-{}-{}-{}-{}-{}-{}-{}-{}-{}-{}-{}-{}-{}-{}-{}-{}-{}-{}-{}-{}-{}-{}-{}-{}-{}-{}-{}-{}} &  \\
  $-1$ & \texttt{-{}-{}-{}-{}-{}-{}-{}-{}-{}-{}-{}-{}-{}-{}-{}-{}-{}-{}-{}-{}-{}-{}-{}-{}-{}-{}-{}-{}-{}-{}-{}-{}} & \texttt{-{}-{}-{}-{}-{}-{}-{}-{}-{}-{}-{}-{}-{}-{}-{}-{}-{}-{}-{}-{}-{}-{}-{}-{}-{}-{}-{}-{}-{}-{}-{}-{}} &  \\
  $0$ & \texttt{-{}-{}-{}-{}-{}-{}-{}-{}-{}-{}-{}-{}-{}-{}-{}-{}-{}-{}-{}-{}-{}-{}-{}-{}-{}-{}-{}-{}-{}-{}-{}-{}} & \texttt{-{}-{}-{}-{}-{}-{}-{}-{}-{}-{}-{}-{}-{}-{}-{}-{}-{}-{}-{}-{}-{}-{}-{}-{}-{}-{}-{}-{}-{}-{}-{}-{}} & \texttt{-{}-{}-{}-{}-{}-{}-{}-{}-{}-{}-{}-{}-{}-{}-{}-{}-{}-{}-{}-{}-{}-{}-{}-{}-{}-{}-{}-{}-{}-{}-{}-{}} \\
  $1$ & \texttt{-{}-{}-{}-{}-{}-{}-{}-{}-{}-{}-{}-{}-{}-{}-{}-{}-{}-{}-{}-{}-{}-{}-{}-{}-{}-{}-{}-{}-{}-{}-{}-{}} & \texttt{-{}-{}-{}-{}-{}-{}-{}-{}-{}-{}-{}-{}-{}-{}-{}-{}-{}-{}-{}-{}-{}-{}-{}-{}-{}-{}-{}-{}-{}-{}-{}-{}} & \texttt{-{}-{}-{}-{}-{}-{}-{}-{}-{}-{}-{}-{}-{}-{}-{}-{}-{}-{}-{}-{}-{}-{}-{}-{}-{}-{}-{}-{}-{}-{}-{}-{}} \\
  $2$ & \texttt{-{}-{}-{}-{}-{}-{}-{}-{}-{}-{}-{}-{}-{}-{}-{}-{}-{}-{}-{}-{}-{}-{}-{}-{}-{}-{}-{}-{}-{}-{}-{}-{}} & \texttt{-{}-{}-{}-{}-{}-{}-{}-{}-{}-{}-{}-{}-{}-{}-{}-{}-{}-{}-{}-{}-{}-{}-{}-{}-{}-{}-{}-{}-{}-{}-{}-{}} & \texttt{-{}-{}-{}-{}-{}-{}-{}-{}-{}-{}-{}-{}-{}-{}-{}-{}-{}-{}-{}-{}-{}-{}-{}-{}-{}-{}-{}-{}-{}-{}-{}-{}} \\
  $3$ & \texttt{-{}-{}-{}-{}-{}-{}-{}-{}-{}-{}-{}-{}-{}-{}-{}-{}-{}-{}-{}-{}-{}-{}-{}-{}-{}-{}-{}-{}-{}-{}-{}-{}} & \texttt{-{}-{}-{}-{}-{}-{}-{}-{}-{}-{}-{}-{}-{}-{}-{}-{}-{}-{}-{}-{}-{}-{}-{}-{}-{}-{}-{}-{}-{}-{}-{}-{}} & \texttt{-{}-{}-{}-{}-{}-{}-{}-{}-{}-{}-{}-{}-{}-{}-{}-{}-{}-{}-{}-{}-{}-{}-{}-{}-{}-{}-{}-{}-{}-{}-{}-{}} \\
  $4$ & \texttt{-{}-{}-{}-{}-{}-{}-{}-{}-{}-{}-{}-{}-{}-{}-{}-{}-{}-{}-{}-{}-{}-{}-{}-{}-{}-{}-{}-{}-{}-{}-{}-{}} & \texttt{-{}-{}-{}-{}-{}-{}-{}-{}-{}-{}-{}-{}-{}-{}-{}-{}-{}-{}-{}-{}-{}-{}-{}-{}-{}-{}-{}-{}-{}-{}-{}-{}} & \texttt{-{}-{}-{}-{}-{}-{}-{}-{}-{}-{}-{}-{}-{}-{}-{}-{}-{}-{}-{}-{}-{}-{}-{}-{}-{}-{}-{}-{}-{}-{}-{}-{}} \\
  $5$ & \texttt{-{}-{}-{}-{}-{}-{}-{}-{}-{}-{}-{}-{}-{}-{}-{}-{}-{}-{}-{}-{}-{}-{}-{}-{}-{}-{}-{}-{}-{}-{}-{}-{}} & \texttt{-{}-{}-{}-{}-{}-{}-{}-{}-{}-{}-{}-{}-{}-{}-{}-{}-{}-{}-{}-{}-{}-{}-{}-{}-{}-{}-{}-{}-{}-{}-{}-{}} & \texttt{-{}-{}-{}-{}-{}-{}-{}-{}-{}-{}-{}-{}-{}-{}-{}-{}-{}-{}-{}-{}-{}-{}-{}-{}-{}-{}-{}-{}-{}-{}-{}-{}} \\
  $6$ & \texttt{-{}-{}-{}-{}-{}-{}-{}-{}-{}-{}-{}-{}-{}-{}-{}-{}-{}-{}-{}-{}-{}-{}-{}-{}-{}-{}-{}-{}-{}-{}-{}-{}} & \texttt{-{}-{}-{}-{}-{}-{}-{}-{}-{}-{}-{}-{}-{}-{}-{}-{}-{}-{}-{}-{}-{}-{}-{}-{}-{}-{}-{}-{}-{}-{}-{}-{}} & \texttt{-{}-{}-{}-{}-{}-{}-{}-{}-{}-{}-{}-{}-{}-{}-{}-{}-{}-{}-{}-{}-{}-{}-{}-{}-{}-{}-{}-{}-{}-{}-{}-{}} \\
  $7$ & \texttt{-{}-{}-{}-{}-{}-{}-{}-{}-{}-{}-{}-{}-{}-{}-{}-{}-{}-{}-{}-{}-{}-{}-{}-{}-{}-{}-{}-{}-{}-{}-{}-{}} & \texttt{-{}-{}-{}-{}-{}-{}-{}-{}-{}-{}-{}-{}-{}-{}-{}-{}-{}-{}-{}-{}-{}-{}-{}-{}-{}-{}-{}-{}-{}-{}-{}-{}} & \texttt{-{}-{}-{}-{}-{}-{}-{}-{}-{}-{}-{}-{}-{}-{}-{}-{}-{}-{}-{}-{}-{}-{}-{}-{}-{}-{}-{}-{}-{}-{}-{}-{}} \\
  $8$ & \texttt{?{}?{}?{}?{}?{}?{}?{}?{}?{}?{}?{}?{}?{}?{}?{}?{}?{}?{}?{}?{}?{}?{}?{}?{}?{}?{}?{}?{}?{}?{}?{}?{}} & \texttt{?{}?{}?{}?{}?{}?{}?{}?{}?{}?{}?{}?{}?{}?{}?{}?{}?{}?{}?{}?{}?{}?{}?{}?{}?{}?{}?{}?{}?{}?{}?{}?{}} & \texttt{x{}?{}?{}?{}?{}?{}?{}?{}?{}?{}?{}?{}?{}?{}?{}?{}?{}?{}?{}?{}?{}?{}?{}?{}?{}?{}?{}?{}?{}?{}?{}?{}} \\
  $9$ & \texttt{?{}?{}?{}?{}?{}?{}?{}?{}?{}?{}?{}?{}?{}?{}?{}?{}?{}?{}?{}?{}?{}?{}?{}?{}?{}?{}?{}?{}?{}?{}?{}?{}} & \texttt{?{}?{}?{}?{}?{}?{}?{}?{}?{}?{}?{}?{}?{}?{}?{}?{}?{}?{}?{}?{}?{}?{}?{}?{}?{}?{}?{}?{}?{}?{}?{}?{}} & \texttt{?{}?{}?{}?{}?{}?{}?{}?{}?{}?{}?{}?{}?{}?{}?{}?{}?{}?{}?{}?{}?{}?{}?{}?{}?{}?{}?{}?{}?{}?{}?{}?{}} \\
  $10$ & \texttt{?{}?{}?{}?{}?{}?{}?{}?{}?{}?{}?{}?{}?{}?{}?{}?{}?{}?{}?{}?{}?{}?{}?{}?{}?{}?{}?{}?{}?{}?{}?{}?{}} & \texttt{?{}?{}?{}?{}?{}?{}?{}?{}?{}?{}?{}?{}?{}?{}?{}?{}?{}?{}?{}?{}?{}?{}?{}?{}?{}?{}?{}?{}?{}?{}?{}?{}} & \texttt{-{}-{}-{}-{}-{}-{}-{}-{}-{}-{}-{}-{}-{}-{}-{}-{}-{}-{}-{}-{}-{}-{}-{}-{}-{}-{}-{}-{}-{}-{}-{}-{}} \\
  $11$ & \texttt{-{}-{}-{}-{}-{}-{}-{}-{}-{}-{}-{}-{}-{}-{}-{}-{}-{}-{}-{}-{}-{}-{}-{}-{}-{}-{}-{}-{}-{}-{}-{}-{}} & \texttt{?{}?{}?{}?{}?{}?{}?{}?{}?{}?{}?{}?{}?{}?{}?{}?{}?{}?{}?{}?{}?{}?{}?{}?{}?{}?{}?{}?{}?{}?{}?{}?{}} & \texttt{-{}-{}-{}-{}-{}-{}-{}-{}-{}-{}-{}-{}-{}-{}-{}-{}-{}-{}-{}-{}-{}-{}-{}-{}-{}-{}-{}-{}-{}-{}-{}-{}} \\
  $12$ & \texttt{-{}-{}-{}-{}-{}-{}-{}-{}-{}-{}-{}-{}-{}-{}-{}-{}-{}-{}-{}-{}-{}-{}-{}-{}-{}-{}-{}-{}-{}-{}-{}-{}} & \texttt{?{}?{}?{}?{}?{}?{}?{}?{}?{}?{}?{}?{}?{}?{}?{}?{}?{}?{}?{}?{}?{}?{}?{}?{}?{}?{}?{}?{}?{}?{}?{}?{}} & \texttt{-{}-{}-{}-{}-{}-{}-{}-{}-{}-{}-{}-{}-{}-{}-{}-{}-{}-{}-{}-{}-{}-{}-{}-{}-{}-{}-{}-{}-{}-{}-{}-{}} \\
  $13$ & \texttt{-{}-{}-{}-{}-{}-{}-{}-{}-{}-{}-{}-{}-{}-{}-{}-{}-{}-{}-{}-{}-{}-{}-{}-{}-{}-{}-{}-{}-{}-{}-{}-{}} & \texttt{?{}?{}?{}?{}?{}?{}?{}?{}?{}?{}?{}?{}?{}?{}?{}?{}?{}?{}?{}?{}?{}?{}?{}?{}?{}?{}?{}?{}?{}?{}?{}?{}} & \texttt{?{}?{}?{}?{}?{}?{}?{}?{}?{}?{}?{}?{}?{}?{}?{}?{}?{}?{}?{}?{}?{}?{}?{}?{}?{}?{}?{}?{}?{}?{}?{}?{}} \\
  $14$ & \texttt{-{}-{}-{}-{}-{}-{}-{}-{}-{}-{}-{}-{}-{}-{}-{}-{}-{}-{}-{}-{}-{}-{}-{}-{}-{}-{}-{}-{}-{}-{}-{}-{}} & \texttt{?{}?{}?{}?{}?{}?{}?{}?{}?{}?{}?{}?{}?{}?{}?{}?{}?{}?{}?{}?{}?{}?{}?{}?{}?{}?{}?{}?{}?{}?{}?{}?{}} & \texttt{-{}-{}-{}-{}-{}-{}-{}-{}-{}-{}-{}-{}-{}-{}-{}-{}-{}-{}-{}-{}-{}-{}-{}-{}-{}-{}-{}-{}-{}-{}-{}-{}} \\
  $15$ & \texttt{-{}-{}-{}-{}-{}-{}-{}-{}-{}-{}-{}-{}-{}-{}-{}-{}-{}-{}-{}-{}-{}-{}-{}-{}-{}-{}-{}-{}-{}-{}-{}-{}} & \texttt{-{}-{}-{}-{}-{}-{}-{}-{}-{}-{}-{}-{}-{}-{}-{}-{}-{}-{}-{}-{}-{}-{}-{}-{}-{}-{}-{}-{}-{}-{}-{}-{}} & \texttt{-{}-{}-{}-{}-{}-{}-{}-{}-{}-{}-{}-{}-{}-{}-{}-{}-{}-{}-{}-{}-{}-{}-{}-{}-{}-{}-{}-{}-{}-{}-{}-{}} \\
  $16$ & \texttt{-{}-{}-{}-{}-{}-{}-{}-{}-{}-{}-{}-{}-{}-{}-{}-{}-{}-{}-{}-{}-{}-{}-{}-{}-{}-{}-{}-{}-{}-{}-{}-{}} & \texttt{-{}-{}-{}-{}-{}-{}-{}-{}-{}-{}-{}-{}-{}-{}-{}-{}-{}-{}-{}-{}-{}-{}-{}-{}-{}-{}-{}-{}-{}-{}-{}-{}} & \texttt{?{}?{}?{}?{}?{}?{}?{}?{}?{}?{}?{}?{}?{}?{}?{}?{}?{}?{}?{}?{}?{}?{}?{}?{}?{}?{}?{}?{}?{}?{}?{}?{}} \\
  $17$ & \texttt{-{}-{}-{}-{}-{}-{}-{}-{}-{}-{}-{}-{}-{}-{}-{}-{}-{}-{}-{}-{}-{}-{}-{}-{}-{}-{}-{}-{}-{}-{}-{}-{}} & \texttt{-{}-{}-{}-{}-{}-{}-{}-{}-{}-{}-{}-{}-{}-{}-{}-{}-{}-{}-{}-{}-{}-{}-{}-{}-{}-{}-{}-{}-{}-{}-{}-{}} & \texttt{-{}-{}-{}-{}-{}-{}-{}-{}-{}-{}-{}-{}-{}-{}-{}-{}-{}-{}-{}-{}-{}-{}-{}-{}-{}-{}-{}-{}-{}-{}-{}-{}} \\
  $18$ & \texttt{-{}-{}-{}-{}-{}-{}-{}-{}-{}-{}-{}-{}-{}-{}-{}-{}-{}-{}-{}-{}-{}-{}-{}-{}-{}-{}-{}-{}-{}-{}-{}-{}} & \texttt{-{}-{}-{}-{}-{}-{}-{}-{}-{}-{}-{}-{}-{}-{}-{}-{}-{}-{}-{}-{}-{}-{}-{}-{}-{}-{}-{}-{}-{}-{}-{}-{}} & \texttt{?{}?{}?{}?{}?{}?{}?{}?{}?{}?{}?{}?{}?{}?{}?{}?{}?{}?{}?{}?{}?{}?{}?{}?{}?{}?{}?{}?{}?{}?{}?{}?{}} \\
  $19$ & \texttt{-{}-{}-{}-{}-{}-{}-{}-{}-{}-{}-{}-{}-{}-{}-{}-{}-{}-{}-{}-{}-{}-{}-{}-{}-{}-{}-{}-{}-{}-{}-{}-{}} & \texttt{-{}-{}-{}-{}-{}-{}-{}-{}-{}-{}-{}-{}-{}-{}-{}-{}-{}-{}-{}-{}-{}-{}-{}-{}-{}-{}-{}-{}-{}-{}-{}-{}} & \texttt{-{}-{}-{}-{}-{}-{}-{}-{}-{}-{}-{}-{}-{}-{}-{}-{}-{}-{}-{}-{}-{}-{}-{}-{}-{}-{}-{}-{}-{}-{}-{}-{}} \\
  $20$ & \texttt{-{}-{}-{}-{}-{}-{}-{}-{}-{}-{}-{}-{}-{}-{}-{}-{}-{}-{}-{}-{}-{}-{}-{}-{}-{}-{}-{}-{}-{}-{}-{}-{}} & \texttt{-{}-{}-{}-{}-{}-{}-{}-{}-{}-{}-{}-{}-{}-{}-{}-{}-{}-{}-{}-{}-{}-{}-{}-{}-{}-{}-{}-{}-{}-{}-{}-{}} & \texttt{-{}-{}-{}-{}-{}-{}-{}-{}-{}-{}-{}-{}-{}-{}-{}-{}-{}-{}-{}-{}-{}-{}-{}-{}-{}-{}-{}-{}-{}-{}-{}-{}} \\
  $21$ & \texttt{-{}-{}-{}-{}-{}-{}-{}-{}-{}-{}-{}-{}-{}-{}-{}-{}-{}-{}-{}-{}-{}-{}-{}-{}-{}-{}-{}-{}-{}-{}-{}-{}} & \texttt{-{}-{}-{}-{}-{}-{}-{}-{}-{}-{}-{}-{}-{}-{}-{}-{}-{}-{}-{}-{}-{}-{}-{}-{}-{}-{}-{}-{}-{}-{}-{}-{}} & \texttt{-{}-{}-{}-{}-{}-{}-{}-{}-{}-{}-{}-{}-{}-{}-{}-{}-{}-{}-{}-{}-{}-{}-{}-{}-{}-{}-{}-{}-{}-{}-{}-{}} \\
  $22$ & \texttt{-{}-{}-{}-{}-{}-{}-{}-{}-{}-{}-{}-{}-{}-{}-{}-{}-{}-{}-{}-{}-{}-{}-{}-{}-{}-{}-{}-{}-{}-{}-{}-{}} & \texttt{-{}-{}-{}-{}-{}-{}-{}-{}-{}-{}-{}-{}-{}-{}-{}-{}-{}-{}-{}-{}-{}-{}-{}-{}-{}-{}-{}-{}-{}-{}-{}-{}} & \texttt{-{}-{}-{}-{}-{}-{}-{}-{}-{}-{}-{}-{}-{}-{}-{}-{}-{}-{}-{}-{}-{}-{}-{}-{}-{}-{}-{}-{}-{}-{}-{}-{}} \\
  $23$ & \texttt{-{}-{}-{}-{}-{}-{}-{}-{}-{}-{}-{}-{}-{}-{}-{}-{}-{}-{}-{}-{}-{}-{}-{}-{}-{}-{}-{}-{}-{}-{}-{}-{}} & \texttt{-{}-{}-{}-{}-{}-{}-{}-{}-{}-{}-{}-{}-{}-{}-{}-{}-{}-{}-{}-{}-{}-{}-{}-{}-{}-{}-{}-{}-{}-{}-{}-{}} & \texttt{-{}-{}-{}-{}-{}-{}-{}-{}-{}-{}-{}-{}-{}-{}-{}-{}-{}-{}-{}-{}-{}-{}-{}-{}-{}-{}-{}-{}-{}-{}-{}-{}} \\
  $24$ & \texttt{-{}-{}-{}-{}-{}-{}-{}-{}-{}-{}-{}-{}-{}-{}-{}-{}-{}-{}-{}-{}-{}-{}-{}-{}-{}-{}-{}-{}-{}-{}-{}-{}} & \texttt{-{}-{}-{}-{}-{}-{}-{}-{}-{}-{}-{}-{}-{}-{}-{}-{}-{}-{}-{}-{}-{}-{}-{}-{}-{}-{}-{}-{}-{}-{}-{}-{}} & \texttt{-{}-{}-{}-{}-{}-{}-{}-{}-{}-{}-{}-{}-{}-{}-{}-{}-{}-{}-{}-{}-{}-{}-{}-{}-{}-{}-{}-{}-{}-{}-{}-{}} \\
  $25$ & \texttt{-{}-{}-{}-{}-{}-{}-{}-{}-{}-{}-{}-{}-{}-{}-{}-{}-{}-{}-{}-{}-{}-{}-{}-{}-{}-{}-{}-{}-{}-{}-{}-{}} & \texttt{-{}-{}-{}-{}-{}-{}-{}-{}-{}-{}-{}-{}-{}-{}-{}-{}-{}-{}-{}-{}-{}-{}-{}-{}-{}-{}-{}-{}-{}-{}-{}-{}} & \texttt{-{}-{}-{}-{}-{}-{}-{}-{}-{}-{}-{}-{}-{}-{}-{}-{}-{}-{}-{}-{}-{}-{}-{}-{}-{}-{}-{}-{}-{}-{}-{}-{}} \\
  $26$ & \texttt{-{}-{}-{}-{}-{}-{}-{}-{}-{}-{}-{}-{}-{}-{}-{}-{}-{}-{}-{}-{}-{}-{}-{}-{}-{}-{}-{}-{}-{}-{}-{}-{}} & \texttt{-{}-{}-{}-{}-{}-{}-{}-{}-{}-{}-{}-{}-{}-{}-{}-{}-{}-{}-{}-{}-{}-{}-{}-{}-{}-{}-{}-{}-{}-{}-{}-{}} & \texttt{-{}-{}-{}-{}-{}-{}-{}-{}-{}-{}-{}-{}-{}-{}-{}-{}-{}-{}-{}-{}-{}-{}-{}-{}-{}-{}-{}-{}-{}-{}-{}-{}} \\
  $27$ & \texttt{-{}-{}-{}-{}-{}-{}-{}-{}-{}-{}-{}-{}-{}-{}-{}-{}-{}-{}-{}-{}-{}-{}-{}-{}-{}-{}-{}-{}-{}-{}-{}-{}} & \texttt{-{}-{}-{}-{}-{}-{}-{}-{}-{}-{}-{}-{}-{}-{}-{}-{}-{}-{}-{}-{}-{}-{}-{}-{}-{}-{}-{}-{}-{}-{}-{}-{}} & \texttt{-{}-{}-{}-{}-{}-{}-{}-{}-{}-{}-{}-{}-{}-{}-{}-{}-{}-{}-{}-{}-{}-{}-{}-{}-{}-{}-{}-{}-{}-{}-{}-{}} \\
}

\newcommand{\SPXXV}{
  $-4$ & \texttt{-{}-{}-{}-{}-{}-{}-{}-{}-{}-{}-{}-{}-{}-{}-{}-{}-{}-{}-{}-{}-{}-{}-{}-{}-{}-{}-{}-{}-{}-{}-{}-{}} & \texttt{-{}-{}-{}-{}-{}-{}-{}-{}-{}-{}-{}-{}-{}-{}-{}-{}-{}-{}-{}-{}-{}-{}-{}-{}-{}-{}-{}-{}-{}-{}-{}-{}} &  \\
  $-3$ & \texttt{-{}-{}-{}-{}-{}-{}-{}-{}-{}-{}-{}-{}-{}-{}-{}-{}-{}-{}-{}-{}-{}-{}-{}-{}-{}-{}-{}-{}-{}-{}-{}-{}} & \texttt{-{}-{}-{}-{}-{}-{}-{}-{}-{}-{}-{}-{}-{}-{}-{}-{}-{}-{}-{}-{}-{}-{}-{}-{}-{}-{}-{}-{}-{}-{}-{}-{}} &  \\
  $-2$ & \texttt{-{}-{}-{}-{}-{}-{}-{}-{}-{}-{}-{}-{}-{}-{}-{}-{}-{}-{}-{}-{}-{}-{}-{}-{}-{}-{}-{}-{}-{}-{}-{}-{}} & \texttt{-{}-{}-{}-{}-{}-{}-{}-{}-{}-{}-{}-{}-{}-{}-{}-{}-{}-{}-{}-{}-{}-{}-{}-{}-{}-{}-{}-{}-{}-{}-{}-{}} &  \\
  $-1$ & \texttt{-{}-{}-{}-{}-{}-{}-{}-{}-{}-{}-{}-{}-{}-{}-{}-{}-{}-{}-{}-{}-{}-{}-{}-{}-{}-{}-{}-{}-{}-{}-{}-{}} & \texttt{-{}-{}-{}-{}-{}-{}-{}-{}-{}-{}-{}-{}-{}-{}-{}-{}-{}-{}-{}-{}-{}-{}-{}-{}-{}-{}-{}-{}-{}-{}-{}-{}} &  \\
  $0$ & \texttt{-{}-{}-{}-{}-{}-{}-{}-{}-{}-{}-{}-{}-{}-{}-{}-{}-{}-{}-{}-{}-{}-{}-{}-{}-{}-{}-{}-{}-{}-{}-{}-{}} & \texttt{-{}-{}-{}-{}-{}-{}-{}-{}-{}-{}-{}-{}-{}-{}-{}-{}-{}-{}-{}-{}-{}-{}-{}-{}-{}-{}-{}-{}-{}-{}-{}-{}} & \texttt{-{}-{}-{}-{}-{}-{}-{}-{}-{}-{}-{}-{}-{}-{}-{}-{}-{}-{}-{}-{}-{}-{}-{}-{}-{}-{}-{}-{}-{}-{}-{}-{}} \\
  $1$ & \texttt{-{}-{}-{}-{}-{}-{}-{}-{}-{}-{}-{}-{}-{}-{}-{}-{}-{}-{}-{}-{}-{}-{}-{}-{}-{}-{}-{}-{}-{}-{}-{}-{}} & \texttt{-{}-{}-{}-{}-{}-{}-{}-{}-{}-{}-{}-{}-{}-{}-{}-{}-{}-{}-{}-{}-{}-{}-{}-{}-{}-{}-{}-{}-{}-{}-{}-{}} & \texttt{-{}-{}-{}-{}-{}-{}-{}-{}-{}-{}-{}-{}-{}-{}-{}-{}-{}-{}-{}-{}-{}-{}-{}-{}-{}-{}-{}-{}-{}-{}-{}-{}} \\
  $2$ & \texttt{-{}-{}-{}-{}-{}-{}-{}-{}-{}-{}-{}-{}-{}-{}-{}-{}-{}-{}-{}-{}-{}-{}-{}-{}-{}-{}-{}-{}-{}-{}-{}-{}} & \texttt{-{}-{}-{}-{}-{}-{}-{}-{}-{}-{}-{}-{}-{}-{}-{}-{}-{}-{}-{}-{}-{}-{}-{}-{}-{}-{}-{}-{}-{}-{}-{}-{}} & \texttt{-{}-{}-{}-{}-{}-{}-{}-{}-{}-{}-{}-{}-{}-{}-{}-{}-{}-{}-{}-{}-{}-{}-{}-{}-{}-{}-{}-{}-{}-{}-{}-{}} \\
  $3$ & \texttt{-{}-{}-{}-{}-{}-{}-{}-{}-{}-{}-{}-{}-{}-{}-{}-{}-{}-{}-{}-{}-{}-{}-{}-{}-{}-{}-{}-{}-{}-{}-{}-{}} & \texttt{-{}-{}-{}-{}-{}-{}-{}-{}-{}-{}-{}-{}-{}-{}-{}-{}-{}-{}-{}-{}-{}-{}-{}-{}-{}-{}-{}-{}-{}-{}-{}-{}} & \texttt{-{}-{}-{}-{}-{}-{}-{}-{}-{}-{}-{}-{}-{}-{}-{}-{}-{}-{}-{}-{}-{}-{}-{}-{}-{}-{}-{}-{}-{}-{}-{}-{}} \\
  $4$ & \texttt{-{}-{}-{}-{}-{}-{}-{}-{}-{}-{}-{}-{}-{}-{}-{}-{}-{}-{}-{}-{}-{}-{}-{}-{}-{}-{}-{}-{}-{}-{}-{}-{}} & \texttt{-{}-{}-{}-{}-{}-{}-{}-{}-{}-{}-{}-{}-{}-{}-{}-{}-{}-{}-{}-{}-{}-{}-{}-{}-{}-{}-{}-{}-{}-{}-{}-{}} & \texttt{-{}-{}-{}-{}-{}-{}-{}-{}-{}-{}-{}-{}-{}-{}-{}-{}-{}-{}-{}-{}-{}-{}-{}-{}-{}-{}-{}-{}-{}-{}-{}-{}} \\
  $5$ & \texttt{-{}-{}-{}-{}-{}-{}-{}-{}-{}-{}-{}-{}-{}-{}-{}-{}-{}-{}-{}-{}-{}-{}-{}-{}-{}-{}-{}-{}-{}-{}-{}-{}} & \texttt{-{}-{}-{}-{}-{}-{}-{}-{}-{}-{}-{}-{}-{}-{}-{}-{}-{}-{}-{}-{}-{}-{}-{}-{}-{}-{}-{}-{}-{}-{}-{}-{}} & \texttt{-{}-{}-{}-{}-{}-{}-{}-{}-{}-{}-{}-{}-{}-{}-{}-{}-{}-{}-{}-{}-{}-{}-{}-{}-{}-{}-{}-{}-{}-{}-{}-{}} \\
  $6$ & \texttt{-{}-{}-{}-{}-{}-{}-{}-{}-{}-{}-{}-{}-{}-{}-{}-{}-{}-{}-{}-{}-{}-{}-{}-{}-{}-{}-{}-{}-{}-{}-{}-{}} & \texttt{-{}-{}-{}-{}-{}-{}-{}-{}-{}-{}-{}-{}-{}-{}-{}-{}-{}-{}-{}-{}-{}-{}-{}-{}-{}-{}-{}-{}-{}-{}-{}-{}} & \texttt{-{}-{}-{}-{}-{}-{}-{}-{}-{}-{}-{}-{}-{}-{}-{}-{}-{}-{}-{}-{}-{}-{}-{}-{}-{}-{}-{}-{}-{}-{}-{}-{}} \\
  $7$ & \texttt{-{}-{}-{}-{}-{}-{}-{}-{}-{}-{}-{}-{}-{}-{}-{}-{}-{}-{}-{}-{}-{}-{}-{}-{}-{}-{}-{}-{}-{}-{}-{}-{}} & \texttt{-{}-{}-{}-{}-{}-{}-{}-{}-{}-{}-{}-{}-{}-{}-{}-{}-{}-{}-{}-{}-{}-{}-{}-{}-{}-{}-{}-{}-{}-{}-{}-{}} & \texttt{-{}-{}-{}-{}-{}-{}-{}-{}-{}-{}-{}-{}-{}-{}-{}-{}-{}-{}-{}-{}-{}-{}-{}-{}-{}-{}-{}-{}-{}-{}-{}-{}} \\
  $8$ & \texttt{x{}?{}?{}?{}?{}?{}?{}?{}?{}?{}?{}?{}?{}?{}?{}?{}?{}?{}?{}?{}?{}?{}?{}?{}?{}?{}?{}?{}?{}?{}?{}?{}} & \texttt{?{}?{}?{}?{}?{}?{}?{}?{}?{}?{}?{}?{}?{}?{}?{}?{}?{}?{}?{}?{}?{}?{}?{}?{}?{}?{}?{}?{}?{}?{}?{}?{}} & \texttt{?{}?{}?{}?{}?{}?{}?{}?{}?{}?{}?{}?{}?{}?{}?{}?{}?{}?{}?{}?{}?{}?{}?{}?{}?{}?{}?{}?{}?{}?{}?{}?{}} \\
  $9$ & \texttt{-{}-{}-{}-{}-{}-{}-{}-{}-{}-{}-{}-{}-{}-{}-{}-{}-{}-{}-{}-{}-{}-{}-{}-{}-{}-{}-{}-{}-{}-{}-{}-{}} & \texttt{?{}?{}?{}?{}?{}?{}?{}?{}?{}?{}?{}?{}?{}?{}?{}?{}?{}?{}?{}?{}?{}?{}?{}?{}?{}?{}?{}?{}?{}?{}?{}?{}} & \texttt{?{}?{}?{}?{}?{}?{}?{}?{}?{}?{}?{}?{}?{}?{}?{}?{}?{}?{}?{}?{}?{}?{}?{}?{}?{}?{}?{}?{}?{}?{}?{}?{}} \\
  $10$ & \texttt{-{}-{}-{}-{}-{}-{}-{}-{}-{}-{}-{}-{}-{}-{}-{}-{}-{}-{}-{}-{}-{}-{}-{}-{}-{}-{}-{}-{}-{}-{}-{}-{}} & \texttt{?{}?{}?{}?{}?{}?{}?{}?{}?{}?{}?{}?{}?{}?{}?{}?{}?{}?{}?{}?{}?{}?{}?{}?{}?{}?{}?{}?{}?{}?{}?{}?{}} & \texttt{-{}-{}-{}-{}-{}-{}-{}-{}-{}-{}-{}-{}-{}-{}-{}-{}-{}-{}-{}-{}-{}-{}-{}-{}-{}-{}-{}-{}-{}-{}-{}-{}} \\
  $11$ & \texttt{-{}-{}-{}-{}-{}-{}-{}-{}-{}-{}-{}-{}-{}-{}-{}-{}-{}-{}-{}-{}-{}-{}-{}-{}-{}-{}-{}-{}-{}-{}-{}-{}} & \texttt{?{}?{}?{}?{}?{}?{}?{}?{}?{}?{}?{}?{}?{}?{}?{}?{}?{}?{}?{}?{}?{}?{}?{}?{}?{}?{}?{}?{}?{}?{}?{}?{}} & \texttt{?{}?{}?{}?{}?{}?{}?{}?{}?{}?{}?{}?{}?{}?{}?{}?{}?{}?{}?{}?{}?{}?{}?{}?{}?{}?{}?{}?{}?{}?{}?{}?{}} \\
  $12$ & \texttt{-{}-{}-{}-{}-{}-{}-{}-{}-{}-{}-{}-{}-{}-{}-{}-{}-{}-{}-{}-{}-{}-{}-{}-{}-{}-{}-{}-{}-{}-{}-{}-{}} & \texttt{?{}?{}?{}?{}?{}?{}?{}?{}?{}?{}?{}?{}?{}?{}?{}?{}?{}?{}?{}?{}?{}?{}?{}?{}?{}?{}?{}?{}?{}?{}?{}?{}} & \texttt{-{}-{}-{}-{}-{}-{}-{}-{}-{}-{}-{}-{}-{}-{}-{}-{}-{}-{}-{}-{}-{}-{}-{}-{}-{}-{}-{}-{}-{}-{}-{}-{}} \\
  $13$ & \texttt{-{}-{}-{}-{}-{}-{}-{}-{}-{}-{}-{}-{}-{}-{}-{}-{}-{}-{}-{}-{}-{}-{}-{}-{}-{}-{}-{}-{}-{}-{}-{}-{}} & \texttt{-{}-{}-{}-{}-{}-{}-{}-{}-{}-{}-{}-{}-{}-{}-{}-{}-{}-{}-{}-{}-{}-{}-{}-{}-{}-{}-{}-{}-{}-{}-{}-{}} & \texttt{-{}-{}-{}-{}-{}-{}-{}-{}-{}-{}-{}-{}-{}-{}-{}-{}-{}-{}-{}-{}-{}-{}-{}-{}-{}-{}-{}-{}-{}-{}-{}-{}} \\
  $14$ & \texttt{-{}-{}-{}-{}-{}-{}-{}-{}-{}-{}-{}-{}-{}-{}-{}-{}-{}-{}-{}-{}-{}-{}-{}-{}-{}-{}-{}-{}-{}-{}-{}-{}} & \texttt{-{}-{}-{}-{}-{}-{}-{}-{}-{}-{}-{}-{}-{}-{}-{}-{}-{}-{}-{}-{}-{}-{}-{}-{}-{}-{}-{}-{}-{}-{}-{}-{}} & \texttt{-{}-{}-{}-{}-{}-{}-{}-{}-{}-{}-{}-{}-{}-{}-{}-{}-{}-{}-{}-{}-{}-{}-{}-{}-{}-{}-{}-{}-{}-{}-{}-{}} \\
  $15$ & \texttt{-{}-{}-{}-{}-{}-{}-{}-{}-{}-{}-{}-{}-{}-{}-{}-{}-{}-{}-{}-{}-{}-{}-{}-{}-{}-{}-{}-{}-{}-{}-{}-{}} & \texttt{-{}-{}-{}-{}-{}-{}-{}-{}-{}-{}-{}-{}-{}-{}-{}-{}-{}-{}-{}-{}-{}-{}-{}-{}-{}-{}-{}-{}-{}-{}-{}-{}} & \texttt{-{}-{}-{}-{}-{}-{}-{}-{}-{}-{}-{}-{}-{}-{}-{}-{}-{}-{}-{}-{}-{}-{}-{}-{}-{}-{}-{}-{}-{}-{}-{}-{}} \\
  $16$ & \texttt{-{}-{}-{}-{}-{}-{}-{}-{}-{}-{}-{}-{}-{}-{}-{}-{}-{}-{}-{}-{}-{}-{}-{}-{}-{}-{}-{}-{}-{}-{}-{}-{}} & \texttt{-{}-{}-{}-{}-{}-{}-{}-{}-{}-{}-{}-{}-{}-{}-{}-{}-{}-{}-{}-{}-{}-{}-{}-{}-{}-{}-{}-{}-{}-{}-{}-{}} & \texttt{?{}?{}?{}?{}?{}?{}?{}?{}?{}?{}?{}?{}?{}?{}?{}?{}?{}?{}?{}?{}?{}?{}?{}?{}?{}?{}?{}?{}?{}?{}?{}?{}} \\
  $17$ & \texttt{-{}-{}-{}-{}-{}-{}-{}-{}-{}-{}-{}-{}-{}-{}-{}-{}-{}-{}-{}-{}-{}-{}-{}-{}-{}-{}-{}-{}-{}-{}-{}-{}} & \texttt{-{}-{}-{}-{}-{}-{}-{}-{}-{}-{}-{}-{}-{}-{}-{}-{}-{}-{}-{}-{}-{}-{}-{}-{}-{}-{}-{}-{}-{}-{}-{}-{}} & \texttt{-{}-{}-{}-{}-{}-{}-{}-{}-{}-{}-{}-{}-{}-{}-{}-{}-{}-{}-{}-{}-{}-{}-{}-{}-{}-{}-{}-{}-{}-{}-{}-{}} \\
  $18$ & \texttt{-{}-{}-{}-{}-{}-{}-{}-{}-{}-{}-{}-{}-{}-{}-{}-{}-{}-{}-{}-{}-{}-{}-{}-{}-{}-{}-{}-{}-{}-{}-{}-{}} & \texttt{-{}-{}-{}-{}-{}-{}-{}-{}-{}-{}-{}-{}-{}-{}-{}-{}-{}-{}-{}-{}-{}-{}-{}-{}-{}-{}-{}-{}-{}-{}-{}-{}} & \texttt{-{}-{}-{}-{}-{}-{}-{}-{}-{}-{}-{}-{}-{}-{}-{}-{}-{}-{}-{}-{}-{}-{}-{}-{}-{}-{}-{}-{}-{}-{}-{}-{}} \\
  $19$ & \texttt{-{}-{}-{}-{}-{}-{}-{}-{}-{}-{}-{}-{}-{}-{}-{}-{}-{}-{}-{}-{}-{}-{}-{}-{}-{}-{}-{}-{}-{}-{}-{}-{}} & \texttt{-{}-{}-{}-{}-{}-{}-{}-{}-{}-{}-{}-{}-{}-{}-{}-{}-{}-{}-{}-{}-{}-{}-{}-{}-{}-{}-{}-{}-{}-{}-{}-{}} & \texttt{-{}-{}-{}-{}-{}-{}-{}-{}-{}-{}-{}-{}-{}-{}-{}-{}-{}-{}-{}-{}-{}-{}-{}-{}-{}-{}-{}-{}-{}-{}-{}-{}} \\
  $20$ & \texttt{-{}-{}-{}-{}-{}-{}-{}-{}-{}-{}-{}-{}-{}-{}-{}-{}-{}-{}-{}-{}-{}-{}-{}-{}-{}-{}-{}-{}-{}-{}-{}-{}} & \texttt{-{}-{}-{}-{}-{}-{}-{}-{}-{}-{}-{}-{}-{}-{}-{}-{}-{}-{}-{}-{}-{}-{}-{}-{}-{}-{}-{}-{}-{}-{}-{}-{}} & \texttt{-{}-{}-{}-{}-{}-{}-{}-{}-{}-{}-{}-{}-{}-{}-{}-{}-{}-{}-{}-{}-{}-{}-{}-{}-{}-{}-{}-{}-{}-{}-{}-{}} \\
  $21$ & \texttt{-{}-{}-{}-{}-{}-{}-{}-{}-{}-{}-{}-{}-{}-{}-{}-{}-{}-{}-{}-{}-{}-{}-{}-{}-{}-{}-{}-{}-{}-{}-{}-{}} & \texttt{-{}-{}-{}-{}-{}-{}-{}-{}-{}-{}-{}-{}-{}-{}-{}-{}-{}-{}-{}-{}-{}-{}-{}-{}-{}-{}-{}-{}-{}-{}-{}-{}} & \texttt{-{}-{}-{}-{}-{}-{}-{}-{}-{}-{}-{}-{}-{}-{}-{}-{}-{}-{}-{}-{}-{}-{}-{}-{}-{}-{}-{}-{}-{}-{}-{}-{}} \\
  $22$ & \texttt{-{}-{}-{}-{}-{}-{}-{}-{}-{}-{}-{}-{}-{}-{}-{}-{}-{}-{}-{}-{}-{}-{}-{}-{}-{}-{}-{}-{}-{}-{}-{}-{}} & \texttt{-{}-{}-{}-{}-{}-{}-{}-{}-{}-{}-{}-{}-{}-{}-{}-{}-{}-{}-{}-{}-{}-{}-{}-{}-{}-{}-{}-{}-{}-{}-{}-{}} & \texttt{-{}-{}-{}-{}-{}-{}-{}-{}-{}-{}-{}-{}-{}-{}-{}-{}-{}-{}-{}-{}-{}-{}-{}-{}-{}-{}-{}-{}-{}-{}-{}-{}} \\
  $23$ & \texttt{-{}-{}-{}-{}-{}-{}-{}-{}-{}-{}-{}-{}-{}-{}-{}-{}-{}-{}-{}-{}-{}-{}-{}-{}-{}-{}-{}-{}-{}-{}-{}-{}} & \texttt{-{}-{}-{}-{}-{}-{}-{}-{}-{}-{}-{}-{}-{}-{}-{}-{}-{}-{}-{}-{}-{}-{}-{}-{}-{}-{}-{}-{}-{}-{}-{}-{}} & \texttt{-{}-{}-{}-{}-{}-{}-{}-{}-{}-{}-{}-{}-{}-{}-{}-{}-{}-{}-{}-{}-{}-{}-{}-{}-{}-{}-{}-{}-{}-{}-{}-{}} \\
  $24$ & \texttt{-{}-{}-{}-{}-{}-{}-{}-{}-{}-{}-{}-{}-{}-{}-{}-{}-{}-{}-{}-{}-{}-{}-{}-{}-{}-{}-{}-{}-{}-{}-{}-{}} & \texttt{-{}-{}-{}-{}-{}-{}-{}-{}-{}-{}-{}-{}-{}-{}-{}-{}-{}-{}-{}-{}-{}-{}-{}-{}-{}-{}-{}-{}-{}-{}-{}-{}} & \texttt{-{}-{}-{}-{}-{}-{}-{}-{}-{}-{}-{}-{}-{}-{}-{}-{}-{}-{}-{}-{}-{}-{}-{}-{}-{}-{}-{}-{}-{}-{}-{}-{}} \\
}

\newcommand{\SPXXI}{
  $-4$ & \texttt{-{}-{}-{}-{}-{}-{}-{}-{}-{}-{}-{}-{}-{}-{}-{}-{}-{}-{}-{}-{}-{}-{}-{}-{}-{}-{}-{}-{}-{}-{}-{}-{}} & \texttt{-{}-{}-{}-{}-{}-{}-{}-{}-{}-{}-{}-{}-{}-{}-{}-{}-{}-{}-{}-{}-{}-{}-{}-{}-{}-{}-{}-{}-{}-{}-{}-{}} &  \\
  $-3$ & \texttt{-{}-{}-{}-{}-{}-{}-{}-{}-{}-{}-{}-{}-{}-{}-{}-{}-{}-{}-{}-{}-{}-{}-{}-{}-{}-{}-{}-{}-{}-{}-{}-{}} & \texttt{-{}-{}-{}-{}-{}-{}-{}-{}-{}-{}-{}-{}-{}-{}-{}-{}-{}-{}-{}-{}-{}-{}-{}-{}-{}-{}-{}-{}-{}-{}-{}-{}} &  \\
  $-2$ & \texttt{-{}-{}-{}-{}-{}-{}-{}-{}-{}-{}-{}-{}-{}-{}-{}-{}-{}-{}-{}-{}-{}-{}-{}-{}-{}-{}-{}-{}-{}-{}-{}-{}} & \texttt{-{}-{}-{}-{}-{}-{}-{}-{}-{}-{}-{}-{}-{}-{}-{}-{}-{}-{}-{}-{}-{}-{}-{}-{}-{}-{}-{}-{}-{}-{}-{}-{}} &  \\
  $-1$ & \texttt{-{}-{}-{}-{}-{}-{}-{}-{}-{}-{}-{}-{}-{}-{}-{}-{}-{}-{}-{}-{}-{}-{}-{}-{}-{}-{}-{}-{}-{}-{}-{}-{}} & \texttt{-{}-{}-{}-{}-{}-{}-{}-{}-{}-{}-{}-{}-{}-{}-{}-{}-{}-{}-{}-{}-{}-{}-{}-{}-{}-{}-{}-{}-{}-{}-{}-{}} &  \\
  $0$ & \texttt{-{}-{}-{}-{}-{}-{}-{}-{}-{}-{}-{}-{}-{}-{}-{}-{}-{}-{}-{}-{}-{}-{}-{}-{}-{}-{}-{}-{}-{}-{}-{}-{}} & \texttt{-{}-{}-{}-{}-{}-{}-{}-{}-{}-{}-{}-{}-{}-{}-{}-{}-{}-{}-{}-{}-{}-{}-{}-{}-{}-{}-{}-{}-{}-{}-{}-{}} & \texttt{-{}-{}-{}-{}-{}-{}-{}-{}-{}-{}-{}-{}-{}-{}-{}-{}-{}-{}-{}-{}-{}-{}-{}-{}-{}-{}-{}-{}-{}-{}-{}-{}} \\
  $1$ & \texttt{-{}-{}-{}-{}-{}-{}-{}-{}-{}-{}-{}-{}-{}-{}-{}-{}-{}-{}-{}-{}-{}-{}-{}-{}-{}-{}-{}-{}-{}-{}-{}-{}} & \texttt{-{}-{}-{}-{}-{}-{}-{}-{}-{}-{}-{}-{}-{}-{}-{}-{}-{}-{}-{}-{}-{}-{}-{}-{}-{}-{}-{}-{}-{}-{}-{}-{}} & \texttt{-{}-{}-{}-{}-{}-{}-{}-{}-{}-{}-{}-{}-{}-{}-{}-{}-{}-{}-{}-{}-{}-{}-{}-{}-{}-{}-{}-{}-{}-{}-{}-{}} \\
  $2$ & \texttt{-{}-{}-{}-{}-{}-{}-{}-{}-{}-{}-{}-{}-{}-{}-{}-{}-{}-{}-{}-{}-{}-{}-{}-{}-{}-{}-{}-{}-{}-{}-{}-{}} & \texttt{-{}-{}-{}-{}-{}-{}-{}-{}-{}-{}-{}-{}-{}-{}-{}-{}-{}-{}-{}-{}-{}-{}-{}-{}-{}-{}-{}-{}-{}-{}-{}-{}} & \texttt{-{}-{}-{}-{}-{}-{}-{}-{}-{}-{}-{}-{}-{}-{}-{}-{}-{}-{}-{}-{}-{}-{}-{}-{}-{}-{}-{}-{}-{}-{}-{}-{}} \\
  $3$ & \texttt{-{}-{}-{}-{}-{}-{}-{}-{}-{}-{}-{}-{}-{}-{}-{}-{}-{}-{}-{}-{}-{}-{}-{}-{}-{}-{}-{}-{}-{}-{}-{}-{}} & \texttt{-{}-{}-{}-{}-{}-{}-{}-{}-{}-{}-{}-{}-{}-{}-{}-{}-{}-{}-{}-{}-{}-{}-{}-{}-{}-{}-{}-{}-{}-{}-{}-{}} & \texttt{-{}-{}-{}-{}-{}-{}-{}-{}-{}-{}-{}-{}-{}-{}-{}-{}-{}-{}-{}-{}-{}-{}-{}-{}-{}-{}-{}-{}-{}-{}-{}-{}} \\
  $4$ & \texttt{-{}-{}-{}-{}-{}-{}-{}-{}-{}-{}-{}-{}-{}-{}-{}-{}-{}-{}-{}-{}-{}-{}-{}-{}-{}-{}-{}-{}-{}-{}-{}-{}} & \texttt{-{}-{}-{}-{}-{}-{}-{}-{}-{}-{}-{}-{}-{}-{}-{}-{}-{}-{}-{}-{}-{}-{}-{}-{}-{}-{}-{}-{}-{}-{}-{}-{}} & \texttt{-{}-{}-{}-{}-{}-{}-{}-{}-{}-{}-{}-{}-{}-{}-{}-{}-{}-{}-{}-{}-{}-{}-{}-{}-{}-{}-{}-{}-{}-{}-{}-{}} \\
  $5$ & \texttt{x{}?{}?{}?{}?{}?{}?{}?{}?{}?{}?{}?{}?{}?{}?{}?{}?{}?{}?{}?{}?{}?{}?{}?{}?{}?{}?{}?{}?{}?{}?{}?{}} & \texttt{?{}?{}?{}?{}?{}?{}?{}?{}?{}?{}?{}?{}?{}?{}?{}?{}?{}?{}?{}?{}?{}?{}?{}?{}?{}?{}?{}?{}?{}?{}?{}?{}} & \texttt{?{}?{}?{}?{}?{}?{}?{}?{}?{}?{}?{}?{}?{}?{}?{}?{}?{}?{}?{}?{}?{}?{}?{}?{}?{}?{}?{}?{}?{}?{}?{}?{}} \\
  $6$ & \texttt{-{}-{}-{}-{}-{}-{}-{}-{}-{}-{}-{}-{}-{}-{}-{}-{}-{}-{}-{}-{}-{}-{}-{}-{}-{}-{}-{}-{}-{}-{}-{}-{}} & \texttt{?{}?{}?{}?{}?{}?{}?{}?{}?{}?{}?{}?{}?{}?{}?{}?{}?{}?{}?{}?{}?{}?{}?{}?{}?{}?{}?{}?{}?{}?{}?{}?{}} & \texttt{?{}?{}?{}?{}?{}?{}?{}?{}?{}?{}?{}?{}?{}?{}?{}?{}?{}?{}?{}?{}?{}?{}?{}?{}?{}?{}?{}?{}?{}?{}?{}?{}} \\
  $7$ & \texttt{-{}-{}-{}-{}-{}-{}-{}-{}-{}-{}-{}-{}-{}-{}-{}-{}-{}-{}-{}-{}-{}-{}-{}-{}-{}-{}-{}-{}-{}-{}-{}-{}} & \texttt{?{}?{}?{}?{}?{}?{}?{}?{}?{}?{}?{}?{}?{}?{}?{}?{}?{}?{}?{}?{}?{}?{}?{}?{}?{}?{}?{}?{}?{}?{}?{}?{}} & \texttt{?{}?{}?{}?{}?{}?{}?{}?{}?{}?{}?{}?{}?{}?{}?{}?{}?{}?{}?{}?{}?{}?{}?{}?{}?{}?{}?{}?{}?{}?{}?{}?{}} \\
  $8$ & \texttt{-{}-{}-{}-{}-{}-{}-{}-{}-{}-{}-{}-{}-{}-{}-{}-{}-{}-{}-{}-{}-{}-{}-{}-{}-{}-{}-{}-{}-{}-{}-{}-{}} & \texttt{?{}?{}?{}?{}?{}?{}?{}?{}?{}?{}?{}?{}?{}?{}?{}?{}?{}?{}?{}?{}?{}?{}?{}?{}?{}?{}?{}?{}?{}?{}?{}?{}} & \texttt{?{}?{}?{}?{}?{}?{}?{}?{}?{}?{}?{}?{}?{}?{}?{}?{}?{}?{}?{}?{}?{}?{}?{}?{}?{}?{}?{}?{}?{}?{}?{}?{}} \\
  $9$ & \texttt{-{}-{}-{}-{}-{}-{}-{}-{}-{}-{}-{}-{}-{}-{}-{}-{}-{}-{}-{}-{}-{}-{}-{}-{}-{}-{}-{}-{}-{}-{}-{}-{}} & \texttt{?{}?{}?{}?{}?{}?{}?{}?{}?{}?{}?{}?{}?{}?{}?{}?{}?{}?{}?{}?{}?{}?{}?{}?{}?{}?{}?{}?{}?{}?{}?{}?{}} & \texttt{-{}-{}-{}-{}-{}-{}-{}-{}-{}-{}-{}-{}-{}-{}-{}-{}-{}-{}-{}-{}-{}-{}-{}-{}-{}-{}-{}-{}-{}-{}-{}-{}} \\
  $10$ & \texttt{-{}-{}-{}-{}-{}-{}-{}-{}-{}-{}-{}-{}-{}-{}-{}-{}-{}-{}-{}-{}-{}-{}-{}-{}-{}-{}-{}-{}-{}-{}-{}-{}} & \texttt{-{}-{}-{}-{}-{}-{}-{}-{}-{}-{}-{}-{}-{}-{}-{}-{}-{}-{}-{}-{}-{}-{}-{}-{}-{}-{}-{}-{}-{}-{}-{}-{}} & \texttt{-{}-{}-{}-{}-{}-{}-{}-{}-{}-{}-{}-{}-{}-{}-{}-{}-{}-{}-{}-{}-{}-{}-{}-{}-{}-{}-{}-{}-{}-{}-{}-{}} \\
  $11$ & \texttt{-{}-{}-{}-{}-{}-{}-{}-{}-{}-{}-{}-{}-{}-{}-{}-{}-{}-{}-{}-{}-{}-{}-{}-{}-{}-{}-{}-{}-{}-{}-{}-{}} & \texttt{-{}-{}-{}-{}-{}-{}-{}-{}-{}-{}-{}-{}-{}-{}-{}-{}-{}-{}-{}-{}-{}-{}-{}-{}-{}-{}-{}-{}-{}-{}-{}-{}} & \texttt{-{}-{}-{}-{}-{}-{}-{}-{}-{}-{}-{}-{}-{}-{}-{}-{}-{}-{}-{}-{}-{}-{}-{}-{}-{}-{}-{}-{}-{}-{}-{}-{}} \\
  $12$ & \texttt{-{}-{}-{}-{}-{}-{}-{}-{}-{}-{}-{}-{}-{}-{}-{}-{}-{}-{}-{}-{}-{}-{}-{}-{}-{}-{}-{}-{}-{}-{}-{}-{}} & \texttt{-{}-{}-{}-{}-{}-{}-{}-{}-{}-{}-{}-{}-{}-{}-{}-{}-{}-{}-{}-{}-{}-{}-{}-{}-{}-{}-{}-{}-{}-{}-{}-{}} & \texttt{-{}-{}-{}-{}-{}-{}-{}-{}-{}-{}-{}-{}-{}-{}-{}-{}-{}-{}-{}-{}-{}-{}-{}-{}-{}-{}-{}-{}-{}-{}-{}-{}} \\
  $13$ & \texttt{-{}-{}-{}-{}-{}-{}-{}-{}-{}-{}-{}-{}-{}-{}-{}-{}-{}-{}-{}-{}-{}-{}-{}-{}-{}-{}-{}-{}-{}-{}-{}-{}} & \texttt{-{}-{}-{}-{}-{}-{}-{}-{}-{}-{}-{}-{}-{}-{}-{}-{}-{}-{}-{}-{}-{}-{}-{}-{}-{}-{}-{}-{}-{}-{}-{}-{}} & \texttt{?{}?{}?{}?{}?{}?{}?{}?{}?{}?{}?{}?{}?{}?{}?{}?{}?{}?{}?{}?{}?{}?{}?{}?{}?{}?{}?{}?{}?{}?{}?{}?{}} \\
  $14$ & \texttt{-{}-{}-{}-{}-{}-{}-{}-{}-{}-{}-{}-{}-{}-{}-{}-{}-{}-{}-{}-{}-{}-{}-{}-{}-{}-{}-{}-{}-{}-{}-{}-{}} & \texttt{-{}-{}-{}-{}-{}-{}-{}-{}-{}-{}-{}-{}-{}-{}-{}-{}-{}-{}-{}-{}-{}-{}-{}-{}-{}-{}-{}-{}-{}-{}-{}-{}} & \texttt{-{}-{}-{}-{}-{}-{}-{}-{}-{}-{}-{}-{}-{}-{}-{}-{}-{}-{}-{}-{}-{}-{}-{}-{}-{}-{}-{}-{}-{}-{}-{}-{}} \\
  $15$ & \texttt{-{}-{}-{}-{}-{}-{}-{}-{}-{}-{}-{}-{}-{}-{}-{}-{}-{}-{}-{}-{}-{}-{}-{}-{}-{}-{}-{}-{}-{}-{}-{}-{}} & \texttt{-{}-{}-{}-{}-{}-{}-{}-{}-{}-{}-{}-{}-{}-{}-{}-{}-{}-{}-{}-{}-{}-{}-{}-{}-{}-{}-{}-{}-{}-{}-{}-{}} & \texttt{-{}-{}-{}-{}-{}-{}-{}-{}-{}-{}-{}-{}-{}-{}-{}-{}-{}-{}-{}-{}-{}-{}-{}-{}-{}-{}-{}-{}-{}-{}-{}-{}} \\
  $16$ & \texttt{-{}-{}-{}-{}-{}-{}-{}-{}-{}-{}-{}-{}-{}-{}-{}-{}-{}-{}-{}-{}-{}-{}-{}-{}-{}-{}-{}-{}-{}-{}-{}-{}} & \texttt{-{}-{}-{}-{}-{}-{}-{}-{}-{}-{}-{}-{}-{}-{}-{}-{}-{}-{}-{}-{}-{}-{}-{}-{}-{}-{}-{}-{}-{}-{}-{}-{}} & \texttt{-{}-{}-{}-{}-{}-{}-{}-{}-{}-{}-{}-{}-{}-{}-{}-{}-{}-{}-{}-{}-{}-{}-{}-{}-{}-{}-{}-{}-{}-{}-{}-{}} \\
  $17$ & \texttt{-{}-{}-{}-{}-{}-{}-{}-{}-{}-{}-{}-{}-{}-{}-{}-{}-{}-{}-{}-{}-{}-{}-{}-{}-{}-{}-{}-{}-{}-{}-{}-{}} & \texttt{-{}-{}-{}-{}-{}-{}-{}-{}-{}-{}-{}-{}-{}-{}-{}-{}-{}-{}-{}-{}-{}-{}-{}-{}-{}-{}-{}-{}-{}-{}-{}-{}} & \texttt{-{}-{}-{}-{}-{}-{}-{}-{}-{}-{}-{}-{}-{}-{}-{}-{}-{}-{}-{}-{}-{}-{}-{}-{}-{}-{}-{}-{}-{}-{}-{}-{}} \\
  $18$ & \texttt{-{}-{}-{}-{}-{}-{}-{}-{}-{}-{}-{}-{}-{}-{}-{}-{}-{}-{}-{}-{}-{}-{}-{}-{}-{}-{}-{}-{}-{}-{}-{}-{}} & \texttt{-{}-{}-{}-{}-{}-{}-{}-{}-{}-{}-{}-{}-{}-{}-{}-{}-{}-{}-{}-{}-{}-{}-{}-{}-{}-{}-{}-{}-{}-{}-{}-{}} & \texttt{-{}-{}-{}-{}-{}-{}-{}-{}-{}-{}-{}-{}-{}-{}-{}-{}-{}-{}-{}-{}-{}-{}-{}-{}-{}-{}-{}-{}-{}-{}-{}-{}} \\
  $19$ & \texttt{-{}-{}-{}-{}-{}-{}-{}-{}-{}-{}-{}-{}-{}-{}-{}-{}-{}-{}-{}-{}-{}-{}-{}-{}-{}-{}-{}-{}-{}-{}-{}-{}} & \texttt{-{}-{}-{}-{}-{}-{}-{}-{}-{}-{}-{}-{}-{}-{}-{}-{}-{}-{}-{}-{}-{}-{}-{}-{}-{}-{}-{}-{}-{}-{}-{}-{}} & \texttt{-{}-{}-{}-{}-{}-{}-{}-{}-{}-{}-{}-{}-{}-{}-{}-{}-{}-{}-{}-{}-{}-{}-{}-{}-{}-{}-{}-{}-{}-{}-{}-{}} \\
  $20$ & \texttt{-{}-{}-{}-{}-{}-{}-{}-{}-{}-{}-{}-{}-{}-{}-{}-{}-{}-{}-{}-{}-{}-{}-{}-{}-{}-{}-{}-{}-{}-{}-{}-{}} & \texttt{-{}-{}-{}-{}-{}-{}-{}-{}-{}-{}-{}-{}-{}-{}-{}-{}-{}-{}-{}-{}-{}-{}-{}-{}-{}-{}-{}-{}-{}-{}-{}-{}} & \texttt{-{}-{}-{}-{}-{}-{}-{}-{}-{}-{}-{}-{}-{}-{}-{}-{}-{}-{}-{}-{}-{}-{}-{}-{}-{}-{}-{}-{}-{}-{}-{}-{}} \\
}

\newcommand{\DCXXXVIII}{
  $-4$ & \texttt{00110100001110000001101001111111} & \texttt{11010010101010100101111010000000} &  \\
  $-3$ & \texttt{11011010011101000111110111100111} & \texttt{11000101100100101011101001101010} &  \\
  $-2$ & \texttt{00011110000010100111001111100010} & \texttt{10001000100101111101110110011000} &  \\
  $-1$ & \texttt{10101111111010100010010101100110} & \texttt{00000110111101001100000011011001} &  \\
  $0$ & \texttt{11111110001111000010010101011001} & \texttt{01101011101011110101110100111011} & \texttt{01011011010100000101100011010010} \\
  $1$ & \texttt{01001011000010110110000101101010} & \texttt{11000001100000000001101010100000} & \texttt{10010000000111111000011111111011} \\
  $2$ & \texttt{01100101111011000011000000100010} & \texttt{11100000101101110101010101101101} & \texttt{00100101010010111100111110100010} \\
  $3$ & \texttt{01110101001010010111110000101000} & \texttt{00001101001111101101010100100110} & \texttt{01011111100011010111110111000001} \\
  $4$ & \texttt{11000011100011110110101000111111} & \texttt{00000010101010001100011111011001} & \texttt{11111011000100000101001110111110} \\
  $5$ & \texttt{11001111011111111011000001010111} & \texttt{00011000001010000101111110011101} & \texttt{00000110001000101110000111111000} \\
  $6$ & \texttt{00111101011110100100110110011011} & \texttt{01001100100101100011110010101001} & \texttt{11011010100010000000000111000010} \\
  $7$ & \texttt{\fbox{{\textcolor{red}{n}}{\textcolor{red}{n}}{\textcolor{red}{n}}0{\textcolor{red}{n}}{\textcolor{red}{n}}{\textcolor{red}{n}}{\textcolor{red}{u}}{\textcolor{red}{n}}{\textcolor{red}{n}}{\textcolor{red}{n}}{\textcolor{red}{n}}{\textcolor{red}{u}}01{\textcolor{red}{u}}{\textcolor{red}{u}}{\textcolor{red}{u}}{\textcolor{red}{u}}{\textcolor{red}{u}}{\textcolor{red}{u}}{\textcolor{red}{u}}{\textcolor{red}{u}}0{\textcolor{red}{u}}{\textcolor{red}{n}}000010}} & \texttt{\fbox{010{\textcolor{red}{u}}1{\textcolor{red}{u}}0{\textcolor{red}{n}}{\textcolor{red}{n}}{\textcolor{red}{n}}{\textcolor{red}{n}}0{\textcolor{red}{n}}{\textcolor{red}{u}}{\textcolor{red}{n}}{\textcolor{red}{n}}{\textcolor{red}{u}}{\textcolor{red}{u}}{\textcolor{red}{u}}{\textcolor{red}{u}}{\textcolor{red}{u}}{\textcolor{red}{u}}{\textcolor{red}{u}}00{\textcolor{red}{u}}001010}} & \texttt{\fbox{10{\textcolor{red}{u}}{\textcolor{red}{n}}{\textcolor{red}{u}}{\textcolor{red}{n}}{\textcolor{red}{n}}10{\textcolor{red}{u}}{\textcolor{red}{n}}10{\textcolor{red}{n}}{\textcolor{red}{n}}1110{\textcolor{red}{n}}{\textcolor{red}{u}}{\textcolor{red}{u}}{\textcolor{red}{u}}1{\textcolor{red}{u}}{\textcolor{red}{n}}111011}} \\
  $8$ & \texttt{11001111111110101001001011100001} & \texttt{\fbox{001{\textcolor{red}{n}}{\textcolor{red}{u}}010011001010000000010010010}} & \texttt{\fbox{0101110110110100001{\textcolor{red}{n}}{\textcolor{red}{u}}{\textcolor{red}{u}}{\textcolor{red}{u}}{\textcolor{red}{u}}{\textcolor{red}{u}}{\textcolor{red}{u}}{\textcolor{red}{u}}{\textcolor{red}{u}}{\textcolor{red}{u}}{\textcolor{red}{u}}01}} \\
  $9$ & \texttt{\fbox{01{\textcolor{red}{u}}001{\textcolor{red}{n}}0{\textcolor{red}{n}}0{\textcolor{red}{u}}{\textcolor{red}{n}}{\textcolor{red}{n}}{\textcolor{red}{n}}{\textcolor{red}{n}}{\textcolor{red}{n}}00{\textcolor{red}{n}}0{\textcolor{red}{n}}{\textcolor{red}{u}}{\textcolor{red}{u}}{\textcolor{red}{n}}{\textcolor{red}{u}}{\textcolor{red}{u}}10{\textcolor{red}{n}}{\textcolor{red}{u}}11}} & \texttt{\fbox{00010111011010010111110001101{\textcolor{red}{n}}00}} & \texttt{01101000001110110100001110010001} \\
  $10$ & \texttt{00000000111000001100000100000111} & \texttt{10001011000001111011101010010110} & \texttt{\fbox{11111000{\textcolor{red}{n}}{\textcolor{red}{u}}{\textcolor{red}{u}}{\textcolor{red}{u}}{\textcolor{red}{u}}{\textcolor{red}{u}}{\textcolor{red}{u}}{\textcolor{red}{n}}{\textcolor{red}{u}}010101110111101}} \\
  $11$ & \texttt{11011000101000000101000100000110} & \texttt{\fbox{0{\textcolor{red}{u}}{\textcolor{red}{n}}{\textcolor{red}{n}}{\textcolor{red}{u}}{\textcolor{red}{n}}011{\textcolor{red}{u}}0{\textcolor{red}{n}}0{\textcolor{red}{u}}1110100{\textcolor{red}{n}}0{\textcolor{red}{u}}{\textcolor{red}{n}}1{\textcolor{red}{n}}{\textcolor{red}{u}}{\textcolor{red}{u}}{\textcolor{red}{u}}01}} & \texttt{11101001001010001011100101110110} \\
  $12$ & \texttt{10110000001000101011101010001100} & \texttt{10000000001000100100001001000101} & \texttt{00110110011101011100110001010101} \\
  $13$ & \texttt{01100100110001111111110111111000} & \texttt{\fbox{1{\textcolor{red}{u}}{\textcolor{red}{n}}0{\textcolor{red}{n}}{\textcolor{red}{u}}0{\textcolor{red}{u}}{\textcolor{red}{u}}{\textcolor{red}{u}}{\textcolor{red}{n}}{\textcolor{red}{n}}{\textcolor{red}{n}}{\textcolor{red}{n}}{\textcolor{red}{n}}{\textcolor{red}{n}}00{\textcolor{red}{n}}1{\textcolor{red}{n}}{\textcolor{red}{u}}{\textcolor{red}{u}}0{\textcolor{red}{n}}{\textcolor{red}{u}}100{\textcolor{red}{n}}11}} & \texttt{01101110101111100111100010111110} \\
  $14$ & \texttt{00001110111110110011010111100111} & \texttt{00110110001100111100010010010000} & \texttt{11100011000000110001010100110110} \\
  $15$ & \texttt{\fbox{1100{\textcolor{red}{n}}0101011001{\textcolor{red}{u}}0100111101111101}} & \texttt{\fbox{1111{\textcolor{red}{n}}110000111{\textcolor{red}{u}}{\textcolor{red}{n}}1011101111111100}} & \texttt{\fbox{11000010110111101001000001101{\textcolor{red}{u}}11}} \\
  $16$ & \texttt{\fbox{11011101100011000101110010100{\textcolor{red}{u}}10}} & \texttt{\fbox{{\textcolor{red}{u}}{\textcolor{red}{n}}{\textcolor{red}{n}}{\textcolor{red}{n}}{\textcolor{red}{n}}{\textcolor{red}{n}}{\textcolor{red}{n}}0001111000110011111{\textcolor{red}{u}}0{\textcolor{red}{n}}{\textcolor{red}{u}}10}} & \texttt{10110000011011100011010101111110} \\
  $17$ & \texttt{11101100000110000000111011010101} & \texttt{\fbox{0001111{\textcolor{red}{n}}{\textcolor{red}{u}}{\textcolor{red}{u}}{\textcolor{red}{u}}111{\textcolor{red}{n}}{\textcolor{red}{u}}0{\textcolor{red}{u}}{\textcolor{red}{n}}{\textcolor{red}{n}}1110101010{\textcolor{red}{n}}{\textcolor{red}{u}}}} & \texttt{11010101101101000100100110000001} \\
  $18$ & \texttt{00110001010011100101110110001111} & \texttt{00000110000111110000111110111100} & \texttt{11000000111100000000000101101111} \\
  $19$ & \texttt{01111001001101010100100010010001} & \texttt{\fbox{0010{\textcolor{red}{n}}101{\textcolor{red}{u}}{\textcolor{red}{n}}{\textcolor{red}{n}}{\textcolor{red}{n}}{\textcolor{red}{n}}{\textcolor{red}{n}}{\textcolor{red}{n}}{\textcolor{red}{n}}0001100110110011}} & \texttt{10111010010001001001101110101000} \\
  $20$ & \texttt{00000011110100011000001001011001} & \texttt{\fbox{100110000101110010110{\textcolor{red}{u}}{\textcolor{red}{n}}{\textcolor{red}{n}}{\textcolor{red}{n}}{\textcolor{red}{n}}{\textcolor{red}{n}}{\textcolor{red}{n}}{\textcolor{red}{n}}{\textcolor{red}{n}}00}} & \texttt{10110011001011110000100101010110} \\
  $21$ & \texttt{11000111011100100100001111011010} & \texttt{01101101111111111111000000000001} & \texttt{01100110010010100111100100010011} \\
  $22$ & \texttt{00100011101011111111110111000001} & \texttt{10101011100001110101011111111110} & \texttt{00010100101010100100011010011101} \\
  $23$ & \texttt{10000010100110001101110100111111} & \texttt{11011110110001000111101101110110} & \texttt{\fbox{1100{\textcolor{red}{u}}000011001{\textcolor{red}{n}}{\textcolor{red}{u}}0000110101111000}} \\
  $24$ & \texttt{01000000000101011100011011111110} & \texttt{01010101111111011000110111111001} & \texttt{\fbox{10011110001100111100000011001{\textcolor{red}{n}}10}} \\
  $25$ & \texttt{11001011100110110010110011011100} & \texttt{10011111100000110000101010100110} & \texttt{00011110010111001111111110111100} \\
  $26$ & \texttt{11100101100100000010111000100010} & \texttt{00001101111100100111101000100101} & \texttt{00101001110010011101001110101010} \\
  $27$ & \texttt{11001011111100101100101000101100} & \texttt{00000100001011101100101111001110} & \texttt{01011100001111001111111000110101} \\
  $28$ & \texttt{11100111000010101101101110111100} & \texttt{10110010110111000001011001011100} & \texttt{01011111100011111001000101011010} \\
  $29$ & \texttt{00101001101011001111100101111010} & \texttt{00111110011010011101111100000101} & \texttt{00011001000111111010011001001101} \\
  $30$ & \texttt{00101111100001101001111000010001} & \texttt{00110111000011000101010101011000} & \texttt{01001001101111110100101000011010} \\
  $31$ & \texttt{11011100111011101011001110011111} & \texttt{11011000010000100000001110000101} & \texttt{11101111001011100111000001011100} \\
  $32$ & \texttt{01111100111101110111110011000001} & \texttt{01100101010010000110100010000001} & \texttt{00100110101001001100110001000010} \\
  $33$ & \texttt{00001111011000110010100101101000} & \texttt{01010111011001000010011011011001} & \texttt{10111101011000000100111100101100} \\
  $34$ & \texttt{11111100101001110111010000000000} & \texttt{00001011000010011100011011001010} & \texttt{01111101101001000011001100101100} \\
  $35$ & \texttt{01001001101101001110110010010100} & \texttt{10100110101111011110100000101010} & \texttt{11000001100001101101001000000100} \\
  $36$ & \texttt{00100010011101111111010111110001} & \texttt{10101100010010101101111100001110} & \texttt{00100000011111001000101001001111} \\
  $37$ & \texttt{00100000111101011111010010010001} & \texttt{10001110110111011111111100011111} & \texttt{00001100100101111011010111101000} \\
}

\newcommand{\IV}{{\it IV}}
\newcommand{\madd}{\mathbin{\boxplus}} % Modular addition
\newcommand{\msub}{\mathbin{\boxminus}} % Modular subtraction
\newcommand{\wmdiff}{{\delta}} % Word pair difference (mod 2^{32})
\newcommand{\bdiff}{{\Delta}} % Bit pair difference (mod 2)
\newcommand{\wconds}{{\nabla}} % Word pair conditions
\newcommand{\F}{\mathbb{F}}

\DeclareMathOperator{\IF}{IF}
\DeclareMathOperator{\XOR}{XOR}
\DeclareMathOperator{\MAJ}{MAJ}

\begin{document}

%% THIS COPYRIGHT BLOCK FOR SC^2 2024!
\copyrightyear{2024}
\copyrightclause{Copyright for this paper by its authors.
  Use permitted under Creative Commons License Attribution 4.0 International (CC BY 4.0).}

%% THIS CONFERENCE BLOCK FOR SC^2 2024!
\conference{9th International Workshop on Satisfiability Checking and Symbolic Computation, July 2, 2024, Nancy, France, Collocated with IJCAR 2024}

\title{SHA-256 Collision Attack with Programmatic SAT}

\author[1]{Nahiyan Alamgir}
\address[1]{University of Windsor, Canada}

%\author[2]{Maria Eichlseder}
%\address[2]{Graz University of Technology, Austria}

\author[2]{Saeed Nejati}
\fnmark[1]
\address[2]{Amazon, USA}

\author[1]{Curtis Bright}[
orcid=0000-0002-0462-625X,
email=cbright@uwindsor.ca,
url=https://www.curtisbright.com/]

\fntext[1]{The contribution to this paper does not relate to this author's position at Amazon.}

\begin{abstract}
Cryptographic hash functions play a crucial role in ensuring data security,
generating fixed-length hashes from variable-length inputs. The hash function
SHA-256 is trusted for data security due to its resilience after over twenty
years of intense scrutiny.  One of its critical properties is collision
resistance, meaning that it is infeasible to find two different inputs with the
same hash.  Currently, the best SHA-256 collision attacks use differential
cryptanalysis to find collisions in simplified versions of SHA-256 that are
reduced to have fewer steps, making it feasible to find collisions.
%step-reduced SHA-256 has a lower number of steps than
%the actual SHA-256 to make the problem substaintially easier.

In this paper, we use a satisfiability (SAT) solver as a tool to search for
step-reduced SHA-256 collisions, and dynamically guide the solver with the aid
of a computer algebra system (CAS) used to detect inconsistencies and deduce
information that the solver would otherwise not detect on its own.  Our hybrid
SAT + CAS solver significantly outperformed a pure SAT approach, enabling us to
find collisions in step-reduced SHA-256 with significantly more steps. Using SAT
+ CAS, we find a 38-step collision of SHA-256 with a modified initialization
vector---something first found by a highly sophisticated search tool of Mendel, Nad, and Schl\"affer.
Conversely, a pure SAT approach could find collisions for no more than 28 steps.
However, our work only uses the SAT solver CaDiCaL and its programmatic
interface IPASIR-UP.
%A key advantage in our methodology is the use of an existing
%finely-tuned generic search tool instead of a custom search tool.
%Overall, this
%research aims to bridge SAT solving with cutting-edge cryptanalysis techniques
%used by \citet{mendel2013improving, li2023new} through a programmatic interface.
\end{abstract}

%%
%% Keywords. The author(s) should pick words that accurately describe % the work
%being presented. Separate the keywords with commas.

%%
%% This command processes the author and affiliation and title % information and
%builds the first part of the formatted document.
\maketitle

\section{Introduction}

Cryptographic hash functions play a vital role in information security. They are
widely relied on for data security and integrity. Due to the high reliance on
cryptographic hash functions for security, they have been constantly targeted
for cryptanalysis. Through these cryptanalytic attacks, some hash functions that
were heavily relied on for security purposes have met their end of life. Some
prominent examples include MD5 and SHA-1---hash functions that were compromised
in terms of collision resistance in the works of \citet{wang2005break} and
\citet{stevens2017first} respectively. SHA-1 was published by NIST in
1995~\cite{sha1}. Six years later, NIST has also published a new family of hash
functions called SHA-2.  Soon after, weaknesses were found in
SHA-1~\cite{Wang2005}, and in 2011 NIST formally recommended anyone relying on
SHA-1 for security migrate to other hash functions like SHA-2.  Consequently,
hash functions in the SHA-2 family have become very widely used. For example,
the function SHA-256 is used for transaction signatures and for proof-of-work in
the Bitcoin protocol~\cite{nakamoto2008}.

Despite the arrival of SHA-3~\cite{sha3}, NIST still recommends both the SHA-2
and SHA-3 families.  SHA-2 is attractive for its ease-of-computation while still
being secure to all known attacks---no collision attack has ever been successful
on the full version, despite a large number of attempts and partial results. One
such attack by \citet{mendel2013improving} in 2013 utilized differential
cryptanalysis for SHA-256.  It was inspired by the 2005 work of
\citet{wang2005break} that used a differential attack (involving modular integer
differences) to find MD5 collisions. Mendel et al.\ found collisions for
step-reduced versions of SHA-256 up to 28 steps and a ``semi-free-start''
collision (where the hash function is slightly modified to allow changing some
predefined constants) of SHA-256 up to 38 steps.  These records held for over
ten years and were only broken in the last few months with the announcement of a
31-step SHA-256 collision~\cite{fse2024} and a 39-step semi-free-start (SFS)
SHA-256 collision~\cite{Li2024}. For more details and background, see
Section~\ref{sec:background}.
%--- recently broken by Li et
%al.\ in the rump session of FSE 2024 with a 31-step collision~\cite{fse2024}.
%Mendel et al. held the record of a 38-step SFS collision as well --- also broken
%recently by \citet{li2023new,li2024new} with a 39-step SFS collision.

Traditionally, the best collisions for step-reduced SHA-256 were found using
highly sophisticated tools specifically designed to search for such collisions.
Conversely, another line of research examined using satisfiability (SAT) solvers
to search for collisions in step-reduced SHA-256, but the results of SAT solvers
were not at all competitive with the best custom-written search tools.  For
example, in 2016, \citet{prokop2016differential} successfully used a SAT solver
to find a collision for 24 steps of SHA-256, but was not able to go higher.  In
2019, \citet{nejati2020} pushed this to 25 steps by using a SAT solver that was
tuned to do programmatic propagation specifically for the collision-finding
problem. However, this was still a long way from Mendel et al.'s 28 step
collision or 38 step SFS collision from 2013~\cite{mendel2013improving}.

In our work, we develop a hybrid approach of using a programmatic
SAT solver that uses a computer algebra system (CAS) to provide the SAT solver
information it wouldn't be able to detect on its own---an approach that
has been successful on many other problems recently~\cite{bright2022}.  
We encode the collision-finding problem directly into SAT (see
Section~\ref{sec:encoding}) and then programmatically encode several of the
mathematical constraints exploited by \citet{mendel2013improving} that made
their 2013 search so effective.

%<<<<<<< HEAD
%This paper is structured as follows. Background is covered in
%Section~\ref{sec:background}---specifically the hash functions, differential
%cryptanalysis, and the $\text{SC}^2$ project. The SAT encoding is discussed in
%Section~\ref{sec:encoding}. In Section~\ref{sec:prog_incons_blocking},
%programmatic inconsistency blocking using linear equations and a graph algorithm
%is described. Programmatic propagation with 2 methods crucial for drastic
%performance improvements over regular SAT has been explained in
%Section~\ref{sec:prog_propagation}. In Section~\ref{sec:impl_and_results}, we
%show the results of our experiment and describe our implementation. Finally, we
%conclude at Section~\ref{sec:conclusion}.
%=======
In particular, we are able to detect and block inconsistencies in the solver's
state using programmatic inconsistency blocking (see
Section~\ref{sec:prog_incons_blocking}) and are able to deduce nontrivial
information about the solver's state using programmatic propagation (see
Section~\ref{sec:prog_propagation}). Our ``SAT + CAS'' solver was able to find
several new 38-step SFS SHA-256 collisions, matching the same step count of the
SFS collisions found by \citet{mendel2013improving}, while a pure SAT approach
was not able to go any further than 28 steps---see Section~\ref{sec:results} for
a summary of our results.

%Overall, our work incorporates various differential cryptanalysis techniques
%employed by \citet{mendel2013improving} to enhance a pure SAT approach.  However,
We note that our SAT-based tool is significantly slower than the dedicated
search tool of Mendel et~al.~\cite{mendel2013improving} for finding 38-step SFS
SHA-256 collisions.  However, the novelty of our work is that we show that the
performance of an off-the-shelf SAT solver can be dramatically improved by
exploiting the IPASIR-UP interface. Moreover, the 38-step SFS SHA-256 collisions
that we found are of independent interest as they have additional structure not
present in \citet{mendel2013improving}'s 38-step SFS SHA-256 collisions---an
additional two internal state variables have zero difference (see
Table~\ref{tbl:dc_38}) when compared with \citet{mendel2013improving}'s
collision.
%our goal is not to compete with their dedicated tool in terms of runtime.
%Instead, we aim to show that an augmented SAT solver can discover SFS collisions
%up to the same number of steps as their tool.
% \CB{Will need a forward reference to an explanation of the ``additional
% structure'' somewhere. Giving the differential path in the appendix would be
% good in order to highlight the additional words with a zero difference.}

% for a SHA-256 collision attack. We employ CaDiCaL, a
%state-of-the-art and modern SAT solver, due to its robust programmatic interface
%(IPASIR-UP \cite{fazekas2023ipasir}) and superior performance. CaDiCaL variants
%and hacks are popular for actively participating in the SAT Competition
%\cite{biere2023cadical}. The IPASIR-UP interface allows us to propagate
%differentials, add blocking clauses to avoid inconsistencies, and take control
%of branching.

%Through our augmented SAT solving techniques, we could reach up to 38 steps of
%semi-free-start collisions, while a plain SAT approach went no further than 28.
%Overall, our contribution is taking differential cryptanalysis techniques of the
%best works and creating a framework based on programmatic SAT (with an embedded
%Computer Algebra System). The main advantage of our technique is that we employ
%a SAT solver, an existing tool highly tuned for search, instead of writing a
%dedicated tool for the collision attack.
%>>>>>>> 9d4a4e6 (Update introduction)

\section{Background}\label{sec:background}

%Our work involves differential cryptanalysis of the hash function SHA-256. Our
%approach relies on a SAT solver for finding collisions. Using the SAT solver's
%programmatic interface, we guide it during search. This section gives an
%overview of cryptographic hash functions, the description of SHA-256, some basic
%concepts of differential cryptanalysis, and finally discussing SAT solving and
%programmatic SAT.

In this section we provide background on cryptographic hash functions,
especially SHA-256, and then discuss differential cryptanalysis and the notation
we use in our work (see Section~\ref{sec:differential_cryptanalysis}). We also
briefly discuss SAT solving and summarize previous collision attacks on SHA-256.

\subsection{Cryptographic Hash Functions}

Cryptographic hash functions take an arbitrary-length input and produce a short
fixed-length output that acts as a signature or fingerprint of the input. The
fingerprint is called a hash value and the input is known as a message. Hash
functions are extensively used for data integrity and security. They are
particularly helpful in cases where storing the message would pose a security
threat but a signature is still required for verification, such as a password in
a database.  The hashes of the passwords can be stored instead and each time the
user enters their password, the hashes can be matched for verification.

Hashes can be used for data integrity as well. For example, when some data is
stored or transferred, the integrity can be checked by comparing a known hash
with the hash of the stored data. This integrity check ensures that the data was
preserved without alteration.

Cryptographic hash functions are expected to have three primary characteristics:

\begin{itemize}
  \item Preimage resistance: It's computationally infeasible to find an input,
  $x$, given a hash, $y$, such that $y$ is the hash of $x$.
  \item 2nd preimage resistance: Given an input, $x$, and its hash, $y$, it's
  infeasible to find a different input, $x'$, that produces the same hash $y$.
  \item Collision resistance: It's infeasible to find an input pair, $x$ and
  $x'$ ($x \neq x'$), that both produce the same hash.
\end{itemize}

%If any of these characteristics break, the hash function can no longer be used
%depending on the purpose.

An example of a weak hash function is MD4~\cite{wang2005cryptanalysis} because
it does not have collision resistance---an attacker can easily generate
colliding message pairs.

\subsection{SHA-256}

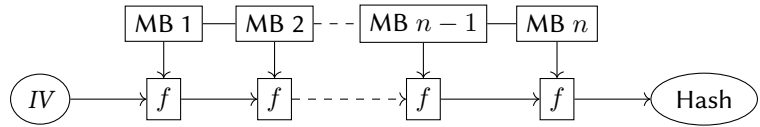
\begin{figure}
  \begin{center}
    \begin{tikzpicture}[]
      % Nodes
      \node [draw, rectangle] (cf1) {$f$};\node [draw, rectangle, right=1 cm of
      cf1] (cf2) {$f$}; \node [draw, rectangle, right=1.5 cm of cf2] (cf3)
      {$f$}; \node [draw, rectangle, right=1.3 cm of cf3] (cf4) {$f$}; \node
      [draw, ellipse, left=1cm of cf1] (iv) {\IV}; \node [draw, rectangle, above
      of=cf1] (mb1) {MB 1}; \node [draw, rectangle, above of=cf2] (mb2) {MB 2};
      \node [draw, rectangle, above of=cf3] (mb3) {MB $n-1$}; \node [draw,
      rectangle, above of=cf4] (mb4) {MB $n$}; \node [draw, ellipse, right=1cm
      of cf4] (hash) {Hash};
      
      % Arrows
      \draw [->] (iv) -- (cf1);
      \draw [-] (mb1) -- (mb2);
      \draw [-, dashed] (mb2) -- (mb3);
      \draw [-] (mb3) -- (mb4);
      \draw [->] (mb1) -- (cf1);
      \draw [->] (mb2) -- (cf2);
      \draw [->] (mb3) -- (cf3);
      \draw [->] (mb4) -- (cf4);
      \draw [->] (cf1) -- (cf2);
      \draw [->, dashed] (cf2) -- (cf3);
      \draw [->] (cf3) -- (cf4);
      \draw [->] (cf4) -- (hash);
    \end{tikzpicture}
  \end{center}
  \caption{SHA-256 processes the input (with padding if needed) into message
  blocks (abbreviated as ``MB''), which are sequentially fed to the compression function, $f$. The
  output of each compression is used as the chaining value in the next
  compression. The compression of the last message block produces the final
  hash. The initial chaining value (\IV) is fixed by the
  specification of SHA-256. The entire method is known as the Merkle--Damgård
  construction, which is popular for building collision-resistant hash
  functions.}\label{fig:MD}
\end{figure}

SHA-256 is a hash function that takes an arbitrary-length input and pads it as
necessary to produce one or many 512-bit message blocks. Afterwards, the message
blocks are processed iteratively to produce a 256-bit hash. Each message block
is processed by a compression function that takes the message block and a
256-bit chaining value as inputs (see Figure~\ref{fig:MD}).

In the compression function, 64 rounds (also called steps) of transformations
are performed to produce a hash. The hash from processing a single message block
is used as the chaining value for the next message block. This means that
altering a message block will lead to cascading changes in the next message
blocks in the sequence. The chaining value for the first message block is set by
the specification~\cite{Dang2015} to a fixed value, known as the
standard $\IV$ (initialization vector).
The hash output is the chaining value produced after applying the
compression function on the last message block.

%\subsubsection{Step-reduced SHA-256}

\begin{figure}
  \begin{center}
    \begin{tikzpicture}[]
      % Nodes
      \node [draw, rectangle] (cf1) {$f_n$};
      \node [draw, ellipse, left=1cm of cf1] (cv) {\textit{CV}};
      \node [draw, rectangle, above of=cf1] (mb1) {Message Block};
      \node [draw, ellipse, right=1cm of cf1] (hash) {Hash};
      
      % Arrows
      \draw [->] (cv) -- (cf1);
      \draw [->] (mb1) -- (cf1);
      \draw [->] (cf1) -- (hash);
    \end{tikzpicture}
  \end{center}
  \caption{A diagram depicting the simplified version of SHA-256 we consider in
  our work. $f_n$ is the step-reduced compression function having $n$ steps. The
  chaining value, \textit{CV}, is arbitrary for semi-free-start (SFS) collisions
  and is not required to match the \IV\ actually used in SHA-256---though a SFS
  colliding message pair is required to have matching
  \textit{CV}s.}\label{fig:simpsha1}
\end{figure}
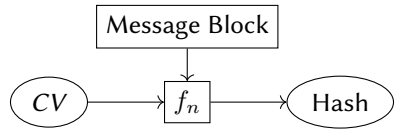

In our work, we focus on a step-reduced version of SHA-256. This means that the
number of rounds/steps in the compression function is reduced to make the
problem easier. Moreover, we only consider messages with a single block of size
512 bits. Because the hash output has 256 bits, there is certainly enough
freedom in the input so that many collisions exist without needing to consider
multiple blocks.
%Moreover, each instance of the hash function has a single
%message block and the chaining value is equal to the standard \IV\ for a
%collision of step-reduced SHA-256.
%\CB{Reviewer 1: it would be nice to give some intuition of how "close" finding an X-step collision is to a generic SHA-256 collision attack in terms of the expected computational effort.}

A relaxed type of collision known as a semi-free-start (SFS) collision allows an
arbitrary initial chaining value \textit{CV}, so long as the \emph{same}
chaining value is used to initialize the hash function for both colliding
messages in the SFS collision (see Figure~\ref{fig:simpsha1}).
In our work, we find SFS collisions for SHA-256 using up to 38 steps of the compression function.
Note that the actual SHA-256 hash function has 64 steps, meaning we are
still very far from finding a true SHA-256 collision (roughly speaking,
as the number of steps increases the collision problem becomes exponentially
more difficult).  The best known collision attacks on SHA-256 are very far from
the full 64 steps, so this provides evidence that SHA-256 is secure.
%A further
%relaxed variant known as a free-start (FS) collision is like an SFS collision
%but with the chaining values not constrained to be equal between the two
%instances.
%
%In our work, we focus on SFS collisions.

\subsubsection{Message Expansion}

SHA-256 performs operations on 32-bit words only. The input message block
consists of 16 such words, $M_i$ for $0 \leq i < 16$, but the compression
function expands the $M_i$ to more words (dependant on $M_0$ to $M_{15}$) to
fill up for the rest of the 64 steps. Altogether there are 64 ``expanded''
message words $W_i$ for $0 \leq i < 64$ defined by
\begin{equation}
  W_i = 
    \begin{cases}
      M_i & \text{for } 0 \leq i < 16\\
      \sigma_1(W_{i-2}) \madd W_{i-7} \madd \sigma_0(W_{i-15}) \madd W_{i-16} & \text{for }
       16 \leq i < 64
    \end{cases}
  \label{eq:upd_w}
\end{equation}
where the functions $\sigma_0$ and $\sigma_1$ are defined as
\begin{align*}
    \sigma_0(X) &= (X \ggg 7) \oplus (X \ggg 18) \oplus (X \gg 3), \text{ and} \\
    \sigma_1(X) &= (X \ggg 17) \oplus (X \ggg 19) \oplus (X \gg 10) .
\end{align*}
Here $\madd$ denotes addition modulo $2^{32}$, $\oplus$ denotes bitwise $\XOR$,
$\gg$ denotes the right shift operator, and $\ggg$ denotes the right
circular shift operator.

\subsubsection{State Update Transformation}\label{sec:stateupdate}

The compression function of SHA-256 takes as input a chaining value and
message block and computes a new chaining value by applying 64 iterations
of a state update procedure.
We describe this state update procedure using equations
similar to those presented by \citet{mendel2011finding}.
%The compression function
%computes the hash using 64 steps starting from the chaining
%value.
The expanded message words $W_i$ are used to compute internal state variables
$T_i$, $E_i$, and $A_i$ through the equations
\begin{align*} 
  &T_i = E_{i-4} \madd \Sigma_1(E_{i-1}) \madd \IF(E_{i-1}, E_{i-2}, E_{i-3})
  \madd K_i \madd W_i, \\
  &E_i = A_{i-4} \madd T_i, \text{ and} \\
  &A_i = T_i \madd \Sigma_0(A_{i-1}) \madd \MAJ(A_{i-1}, A_{i-2}, A_{i-3}) .
\end{align*}
Here the functions $\IF$ and $\MAJ$ are defined on words by applying bitwise the
functions from $\F_2^3$ to $\F_2$
\[
  \IF(x,y,z) = xy + xz + z, \qquad\text{and}\qquad
  \MAJ(x,y,z) = xy + yz + xz, \label{eq:if}
\]
and the linear functions $\Sigma_0$ and $\Sigma_1$ are defined by
\begin{align*} 
  &\Sigma_0(X) = (X \ggg 2) \oplus (X \ggg 13) \oplus (X \ggg 22), \text{ and} \\
  &\Sigma_1(X) = (X \ggg 6) \oplus (X \ggg 11) \oplus (X \ggg 25).
\end{align*}

The chaining value is taken to be $[A_{-4}, \dotsc, A_{-1}, E_{-4},
\dotsc, E_{-1}]$.  In other words, the chaining value sets the initial values of
the state variables $A$ and $E$. For example, $A_{-4}$ will be initialized to
the first 32-bit word of the chaining value while $E_{-1}$ will be initialized
to the last word of the chaining value. $K_i$ is a constant given in SHA-256's
specification and there is one unique constant for each step $i$. The auxiliary
variable $T_i$ is introduced to keep the modular additions from having more than
5 addends.

After the state update transformations, the last four $A$ and $E$ words are
added with the chaining value to produce a new chaining value, which will be the
output of the compression function (and following the final block will be the
output of the hash function).

\subsection{Differential Cryptanalysis}\label{sec:differential_cryptanalysis}

Differential cryptanalysis is a technique that analyzes how the input
differences influence the output differences in, for example, a hash function.
This technique is crucial in collision attacks of hash functions, since we're
interested in studying the diffusion of the input differences to the output
differences such that we get a zero output difference and a non-zero input
difference.

In differential cryptanalysis of hash collisions, we have two hash inputs and
we examine the differences in all the operations until the output for both
inputs.  Usually differences between the values are calculated through $\XOR$
operations, such as $\bdiff x = x \oplus x'$, where $x$ is a single bit,
$x'$ is its counterpart in the second hash instance, and $\bdiff x$ is the
difference of $x$ and $x'$.
%
%The differences in the input and output values of a function or an operation is
%denoted as a differential. For example, the differential $\bdiff x \rightarrow
%\bdiff y$ over a function $f(x)=y$ describes the differences in the input values $(x,
%x')$ over the output values $(y, y')$ of the function $f$.
%
%Hash functions such as SHA-256 operate on bitvectors (also called words) of 32
%bits in length. If a Boolean variable representing a bit in one bitvector and
%its counterpart in the second hash instance form a pair, in general the pair may
In general, a pair $(x,x')$ representing a bit in one bitvector and its counterpart in the second hash instance
may have up to 4 combinations, $\{(0, 0), (0, 1), (1, 0), (1, 1)\}$.
In some cases, some of the four possibilities may be ruled out, though.
The possibilities for $(x,x')$ can be generalized as the differential
conditions~\cite{de2006finding} presented in Table~\ref{tbl:differential_conditions}. For
example, if a pair $(x,x')$ has the possibilities $\{(0, 0), (1, 1)\}$, we
describe it as having the differential condition `\verb|-|',
whereas the differential condition `\verb|x|' describes the possibilities
$\{(0, 1), (1, 0)\}$.

\begin{table}
\caption{This table shows the notation we use for differential conditions in our
study. A `+' indicates whether a specific value pair is possible for $(x,x')$.
For example, `\texttt{?}' indicates that the variables $x$ and $x'$ can take any
value, `\texttt{x}' indicates the variables $x$ and $x'$ have distinct values,
and `\texttt{-}' represents equal values. In the rest of the conditions, the
exact values of the variables $x$ and $x'$ are known.}
\label{tbl:differential_conditions}
\begin{tabular}{clllllll}
    \toprule
  ($x$, $x'$) & \texttt{0} & \texttt{u} & \texttt{n} & \texttt{1} & \texttt{x} &
  \texttt{-} & \texttt{?} \\
    \midrule
  (0, 0)      & + &   &   &   &   & + & +  \\ 
  (1, 0)      &   & + &   &   & + &   & +  \\ 
  (0, 1)      &   &   & + &   & + &   & +  \\ 
  (1, 1)      &   &   &   & + &   & + & +  \\
    \bottomrule
\end{tabular}
\end{table}

For convenience, the differential conditions of a pair of words
$(A, A')$ can be collectively described in a vector $\wconds A = [c_{n}c_{n-1} \cdots
c_1c_0]$, where $c_i$ is the differential condition of the $i$th bit pair $(a_i,
a'_i)$ with $A=[a_{n-1}\cdots a_0]$ and $A'=[a'_{n-1}\cdots a'_0]$.
%([a_{n-1}\cdots a_0],[a'_{n-1}\cdots a'_0])

%As such, the differential of the input message differences (represented by differential
%conditions of words $\wconds M_0$, $\dotsc$, $\wconds M_{15}$),
The differential over a function $f(X)=Y$ where $X$ and $Y$ are bitvectors
%$\wconds X$, and output differences, $\wconds Y$, over a
is denoted $\wconds X \rightarrow \wconds Y$. On a high level, we want $f$ to be
the hash function while $\wconds X$ and $\wconds Y$ are the input and output
differences represented by differential conditions. In practice, analyzing this
differential alone is not helpful as it contains too little information. We want
to study all the operations in between as well---chaining the operations in
SHA-256 together as a series of steps starting from the input to the output. If
we represent the differences in an operation's input and output values as a
differential, we can represent the 2 hash function instances as a chain of
differentials. This chain of differentials is called the differential path and
analyzing this path shows how the differences propagate from the input
differences all the way to the output differences, which is essential for
finding collisions.

%Analyzing every component of the differential path helps us derive more
%information through various techniques, such as propagation of differential
%conditions, as well as other conditions such as the two-bit conditions
%explained in Section~\ref{sec:prog_incons_blocking}. All of the information is
%crucial in shaping the differential path such that it leads to a collision.

\subsection{Boolean Satisfiability (SAT)}

Boolean satisfiability (SAT) solving involves searching for a solution of a
Boolean formula (an assignment of the variables that makes the formula true).
The tools designed for this
purpose are called SAT solvers.
%SAT solving is NP-complete, which means that the
%solution to a SAT problem can be verified in polynomial time, but currently
%there is no known way to solve the problem in polynomial time.
Even though all
known algorithms for SAT solving run in exponential time in the worst case, in
practice many problems can be solved by modern SAT solvers in a reasonable
amount of time. In fact, SAT solvers are so effective that in practice there are
problems unrelated to logic that are most effectively solved by reducing them to
SAT and calling a SAT solver.

The beauty of SAT solving lies in its generic nature, which means that it can be
applied to any domain as long as the problem can be encoded into a Boolean
formula. This also allows solvers to be tuned for performance independently of a
specific problem.  Modern SAT solvers can be surprisingly effective at solving
SAT problems by incorporating sophisticated techniques like conflict analysis and clause learning,
clever branching heuristics, and simplification~\cite{biere2009handbook}.
This combination allows them to be highly
potent at general-purpose search.

\paragraph{The $\mathbf{\text{SC}^2}$ Project.}

SAT solving, ever since its inception, has been found to be useful for
satisfiability checking---determining whether a logical formula is satisfiable.
%For example, \citet{bright2021sat} provided a nonexistence
%certificate of a projective plane of order ten through SAT.
%
On the other side, Computer Algebra Systems (CASs) include algorithms that are
efficient at solving mathematical problems. However, many problems exist that
involves both satisfiability checking and symbolic computation, and bridging the
two fields was proposed in 2015 in the work of \citet{abraham2015building} and
\citet{Zulkoski2015}. Shortly afterwards, the $\text{SC}^2$
project~\cite{abraham2017satisfiability} was initiated to support the joint community.  
Since then, a wide variety of problems have been tackled using SC$^2$
techniques, from circuit verification~\cite{Kaufmann2019} to knot
theory~\cite{Lisitsa2017} to quantifier elimination and cylindrical algebraic
decomposition~\cite{Brown2017}, and factoring integers with known bits~\cite{ajani2024}.
See England's survey~\cite{England2022a} for an
overview of many other examples.
%Prominent works include that of
%\citet{kaufmann2019verifying} where the joint approach was used for verifying
%integer multiplifers in an automated way. Another remarkable work is by
%\citet{brown202072} where they present a partial solver (that doesn't
%necessarily solve the problem) that can determine the satisfiability quickly by
%combining SMT/SAT solving with symbolic computation. \citet{bright2022} also
%highlighted some prime cases where Computer Algebra Systems were employed to
%solve subproblems that weren't efficiently solved through SAT solving.

\paragraph{Programmatic SAT.}

%Often times, the algorithms housed in CASs need to be used during SAT solving. A
%common way is to embed a CAS in a SAT solver and modify the internals to
%interleave --- for example, a CAS can solve a subproblem of a SAT problem and
%add learned clauses to the SAT solver on the fly. The algorithms can be written
%programmatically (for example, using high-level languages like C/C++) while the
%SAT problem is written as a Boolean formula --- together called programmatic
%SAT.

Programmatic SAT involves injecting code into the solver to aid the solver and
solve a problem more efficiently than it otherwise could~\cite{Ganesh2012}. It's
useful especially when aspects of the problem are difficult to express in
conjunctive normal form, the typical input of SAT
solvers~\cite{bright2016mathcheck2}.

Programmatic SAT is usually domain-specific and combines the powerful techniques
of search possessed by SAT solvers with the most efficient high-level algorithms
and analysis tools for the problem.  Some modern SAT solvers can be customized
in a programmatic way through built-in interfaces like
IPASIR-UP~\cite{fazekas2023ipasir}. The solver may be aided through the
following ways during search:
\begin{itemize}
  \item \textbf{External propagation:} Assigning variables derived through
  high-level deduction.
  \item \textbf{External decisions:} Making decisions on picking important
  unassigned variables and guessing their values through high-level analysis.
  \item \textbf{External learning:} Injecting learned clauses during conflicts
  detected through the high-level conflict analysis.
\end{itemize}

%Overall, the external routines act as an oracle to the SAT solver from where it
%gets information that isn't available in the Boolean formula.

\subsection{Previous Work}\label{sec:previous_work}

Cryptographic hash functions such as MD4, MD5, SHA-1, etc.\ have been extensively
relied on for information security for many years.
However, \citet{wang2005cryptanalysis} devised an efficient method in 2005 for
finding MD4 collisions with probability from $2^{-6}$ to $2^{-2}$
using at most 256 MD4 hash operations.
%--- contrary
%to Dobbertin's attack equivalent to $2^{20}$ computations in 1996.
Wang et al.\ also proposed an attack on MD5~\cite{wang2005break} for finding
collisions within 15--60 minutes of computational time in the same year.  Also
in 2005, \citet{wang2005finding} presented a collision attack on SHA-1 using at
most $2^{69}$ SHA-1 hash computations, resulting in
a SHA-1 collision found in 2017~\cite{Stevens2017}.

The SHA-2 family of hash functions, however, survived these remarkable attacks,
likely due to their relatively complex design with message expansion. One of the
earliest attacks on SHA-256 and its family members was in 2003 by
\citet{gilbert2003security}.
%The work also pointed out that attacks on Merkle--Damgård-type
%hash functions do not apply to SHA-2 due to the way it was designed.
%
In FSE 2006, \citet{mendel2006analysis} reported that the message expansion of
the SHA-2 family of hash functions was one of the key points for their increased
collision resilience over SHA-1. To tackle this, Mendel~et~al.\ applied a
message modification technique and reached an 18-step collision for SHA-256. In
INDOCRYPT 2008, Sanadhya and Sarkar~\cite{sanadhya2008new} presented collisions up to 24 steps of
SHA-256 and SHA-512, making improvements over the work of
\citet{nikolic2008collisions} that presented collisions up to 21 steps of
SHA-256 at FSE 2008.

In ASIACRYPT 2011, \citet{mendel2011finding} revealed a collision for 27-step
SHA-256 and a semi-free-start (SFS) collision for 32 steps. They automated the
search with a domain-specific tool that searches for differential
characteristics for SHA-256. The tool utilizes propagation, analysis of the bit
constraints, clever branching on the most constraints bits, and contradiction
detection in the differential characteristics.

In EUROCRYPT 2013, \citet{mendel2013improving} came back with another
breakthrough---a 28-step collision of SHA-256 along with a 38-step SFS
collision. They further improved the automatic search tool that finds
differential characteristics. The improvements included local collisions over a
larger number of steps and improved decision/branching heuristics
over~\cite{mendel2011finding}. % --- fine-tuning the ordering of decisions to specific
%types of variables. Their work also included a 31-step SFS collision but with
%the last 5 words of the chaining values fixed --- closer to an actual collision.

%In 2016, \citet{prokop2016differential} demonstrated finding a 24-step collision
%of SHA-256 using a SAT solver --- a generic search tool, contrary to custom
%search tools used in the leading works. Prokop's work bridged the gap between
%SAT solving and differential cryptanalysis with some of Mendel et al.'s
%approaches.
%
%In CASCON 2019, \citet{nejati2020} came up with a new SAT attack on SHA-256 ---
%it featured an improved SAT encoding (over Prokop's encoding) and also made use
%of a programmatic SAT approach for differential cryptanalysis. Their work was
%also inspired by Mendel et al.'s advancements in cryptanalysis, and found
%collision for 25-step SHA-256. The programmatic SAT approach involved adding a
%programmatic routine on top of MapleSAT \cite{DBLP:conf/sat/LiangGPC16} for
%propagating differentials. In contrast to Prokop's usage of 6 differential
%conditions, Nejati et al. extended it to 16 differential conditions, just like
%that in Mendel et al.'s work.

Very recently, in the rump session of FSE 2024, \citet{fse2024} announced
a 31-step collision of SHA-256, and in a EUROCRYPT 2024 paper
found a 39-step SFS collision~\cite{Li2024}.
%These approaches made use of an SMT/SAT solver
%finding SFS collisions of SHA-256 up to 39 steps --- breaking the EUROCRYPT 2013
%record that was held for around 11 years.
These works also made an advancement in
cryptanalysis with SAT solving, searching for characteristics by
controlling the sparsity (number of variables with no difference).

%In summary, the best known SHA-256 attacks found SFS collisions till 39 steps
%and collisions till 31 steps.
The progress of the attacks on SHA-256 is presented in
Table~\ref{tbl:collisions_progress}.

\begin{table}
  \centering
  \caption{Progress of step-reduced SHA-256 collision attacks (including SFS
    collisions) from 2006 to 2024. The entries in the table indicate the number
    of steps for which the collisions (or SFS collisions) were found.}
  \label{tbl:collisions_progress}
  \begin{tabular}{cccc}
    \toprule
    \textbf{Publication Year} & \textbf{Author} & \textbf{Collision} &
    \textbf{SFS Collision} \\
    \midrule
    2006 & \citet{mendel2006analysis} & 18 & \texttt{-} \\
    2008 & \citet{sanadhya2008new} & 24 & \texttt{-} \\
    2011 & \citet{mendel2011finding} & 27 & 32 \\
    2013 & \citet{mendel2013improving} & 28 & 38 \\
    2024 & \citet{fse2024,Li2024} & 31 & 39 \\
    \bottomrule
  \end{tabular}
\end{table}

\section{The SAT Encoding}\label{sec:encoding}

Our problem is to find two different messages 
whose hashes match, and to do this we use SAT solvers as a search tool for the collision
attack. %SAT solvers take a Boolean formula and try to find variables that
%satisfy the formula (making it evaluate to true).
Our SAT formula contains variables encoding two messages (each containing one 512-bit message block)
that we want to collide after applying SHA-256.  For
each block, the formula includes an $n$-step compression function taking a
512-bit message block and a 256-bit chaining value, and from them computes a 256-bit hash.
The number of steps/rounds, $n$, is adjusted to generate a step-reduced version of
SHA-256.

Encoding the compression function includes bitwise Boolean functions such as
$\IF$ and $\MAJ$. The other functions, $\sigma_0$, $\sigma_1$, $\Sigma_0$, and
$\Sigma_1$, boil down to 3-operand $\XOR$ functions after circular rotations and
shifts. For each 3-bit $\XOR$ $a \oplus b \oplus c$, our encoding produces $x
\leftrightarrow a \oplus b \oplus c$ (where $x$ is a new auxiliary variable) using
$2^3=8$ clauses. A new variable is introduced for every gate in the
circuit similar to how the Tseitin transformation is
performed~\cite{Tseitin1983}.

The 32-bit modular addition is encoded as bitslices, where each bitslice
involves at most 7 addends (including carries) and a 3-bit output (a high carry,
a low carry, and a sum). The addition encoding is taken from the work of
\citet{nejati2020}, which used the Espresso logic minimizer~\cite{Rudell1987}.

On top of the two hash function instances, we have the differential
cryptanalysis layer. Each Boolean variable in one instance, say $x$, has its
counterpart $x'$ in the other instance. For the analysis of the differences as
per differential cryptanalysis, we encode the bitwise differences as $\bdiff x
\leftrightarrow x \oplus x'$ (following \citet{nejati2020}) where $\bdiff x \in
\{0, 1\}$ is a new auxiliary variable. Each triple ($x$, $x'$, $\bdiff x$)
defines a differential condition. For example, $(x, x', 1)$ defines an
\texttt{x} while $(1, 0, \bdiff x)$ defines a \texttt{u} (see
Table~\ref{tbl:differential_conditions} for the complete list of differential
conditions).

A naive way to constrain collisions is to have zero differences in the hash pair
while maintaining at least one difference in the message pair. However, we want
to analyze all the differences between the two hash instances, especially the
state update as well as the auxiliary variables, to capture as much information
as possible. Thus, we follow the idea of a local collision as presented in the
works of \citet{mendel2011finding, mendel2013improving} and many others.

To induce a local collision, we constrain the differential conditions in the
state update variables, $A$ and $E$, along with the message words, $W$. The
encoding includes clauses for constraining of the conditions as such, and is
called the \textbf{starting point} of the differential path.
%(more on subsection
%\ref{sec:differential_cryptanalysis}).
For example, if a differential condition on the variable $x$ is constrained to
be a `\texttt{-}', we add the unit clause $\lnot\bdiff x$ to set the difference
to be zero (and thus $x = x'$). The explicit starting points we used in our work are
given in the appendix (Tables~\ref{tbl:sp_21}--\ref{tbl:sp_38}).

Any solution found by the solver will be within the confinements set by the
starting point. This reduction of search space is found to very beneficial and
the possibility and time required for finding collisions highly depends on a
well-crafted starting point.

To make the base problem easier, we also add clauses for the propagation of
common differentials, especially ones with \texttt{-} and \texttt{x}. For
example, the encoding has a clause for propagation of $\verb|[xx-]| \rightarrow
\verb|[-]|$ for the $\XOR$ function, encoding that when $x$ is the auxiliary
variable for $a \oplus b \oplus c$ we have $(\Delta a\land \Delta
b\land\lnot\Delta c)\rightarrow\lnot\Delta x$. Such helpful clauses are present
for all the operations $\XOR$, $\MAJ$, $\IF$, and the modular addition of words
from equation~\eqref{eq:upd_w} and the state update equations of
Section~\ref{sec:stateupdate}.

%As we are searching for semi-free-start collisions, we do not set the chaining
%value but we do constrain the chaining values of the 2 hash instances to be
%equal. This is part of the starting point's characteristics, which also
%constrains the hashes to be equal.

\section{Programmatic Inconsistency Blocking} \label{sec:prog_incons_blocking}

%\begin{figure}
%  \begin{center}
%    \begin{tikzpicture}[
%      node distance=2cm,
%    ]
%      % Nodes
%      \node[circle, draw] (1) {a};
%      \node[circle, draw, right=of 1] (2) {b};
%      \node[circle, draw, below=of 2] (3) {c};
%      \node[circle, draw, below=of 1] (4) {d};
%    
%      % Edges
%      \draw (1) -- node[midway, above] {$=$} (2);
%      \draw (2) -- node[midway, right] {$=$} (3);
%      \draw (3) -- node[midway, below] {$=$} (4);
%      \draw (4) -- node[midway, left] {$\neq$} (1);
%    \end{tikzpicture}
%  \end{center}
%  \caption{The graph shows linear equations between the Boolean variables $a$,
%  $b$, $c$, $d$. The set of equations is inconsistent because $a=b=c=d$ but $a
%  \neq d$. This is an example of an inconsistent cycle.}
%\end{figure}

As discussed in Section~\ref{sec:differential_cryptanalysis}, analyzing
differential paths is essential in cryptanalysis. There are cases when a
differential path has inconsistencies. In other words, parts of the differential
path define a relation contradicting a relation defined by other parts of the
differential path. For example, if we can derive the conditions $a = b$ and $a
\neq b$ from differential conditions in the same path, then there certainly
cannot exist any message pairs conforming to that path. During solving, it's
crucial to analyze the current path for such inconsistencies and block them as
early as possible to prevent the solver from exploring paths that are
inconsistent.

The idea of looking for and blocking inconsistencies in the SHA-256 collision
attack was utilized by \citet{mendel2011finding}. They described having linear
equations relating two Boolean variables in SHA-256's state. Each of these
equations can be derived from bitsliced differentials of bitwise functions and
modular addition. Such relations can lead to conditions on the equality or
inequality of two variables. \citet{mendel2011finding, mendel2013improving}
refers to these conditions as ``two-bit conditions''.

Two-bit conditions can be derived from bitsliced differentials of bitwise
functions and addition operations. To deduce the two-bit conditions from a
bitsliced differential, we enumerate all the possibilities and look for a
pattern. As an %\CB{Updated the notation.  But is it accurate to call this a
%``bitsliced'' differential?} \NA{I think you're right since we defined this as a
%3-bit operation.}
example, consider the differential $\wconds[x_2 x_1 x_0]
\rightarrow \wconds [y_0]$ of the $\XOR$ operation $x_0\oplus x_1\oplus
x_2=y_0$.  If the differential is specifically $\verb|[-0-]| \rightarrow
\verb|[0]|$, it means that $(x_2, x_0) \in \{(0, 0), (1, 1)\}$ giving us the
two-bit condition $x_2 = x_0$.

In practice, two-bit conditions are significantly more common in the bitsliced
differentials of bitwise functions than that of addition operations. %\CB{The
%first sentence was basically already said in the previous paragraph.} \NA{I
%rewrote some parts of this paragraph now. I meant that two-bit conditions are
%much more common in differentials of bitwise functions, which wasn't stated in
%the previous paragraph.}
Thus, in our experiments, we only computed the two-bit
conditions of these bitwise functions to reduce computational costs.
Additionally, there are two-bit conditions involving Boolean variables other
than that of $A$, $E$, and $W$. For example, two-bit conditions often involve
the output variables of bitwise functions along with other auxiliary variables.
However, we did not find it beneficial to address inconsistencies involving
two-bit conditions of these auxiliary variables. %\CB{``We did not find it
%beneficial'' means it didn't help the solver? Did Mendel et al. mention this as
%well?} \NA{We found increased runtimes with the extra two-bit conditions, and
%none of Mendel et al.'s two-bit conditions (from their paper) involved any of
%the auxiliary variables (including their examples and the lists of two-bit
%conditions).}
As a result, we focused solely on the two-bit conditions involving
the primary variables to block inconsistencies. 

The two-bit conditions are expressed in the form $x \oplus y = z$ where $x$ and
$y$ are variables in the differential path and $z \in \{0, 1\}$.  For example,
the two-bit condition $x \neq y$ gives the equation $x \oplus y = 1$. The set of
these linear equations often lead to inconsistencies that are non-trivial.  For
example, if $a = b$, $b = c$, and $c \neq a$, we have a contradiction involving
the 3 two-bit conditions and can be visualized as a cycle $a=b=c\neq a$.
%--- can be visualized as a
%cycle 
%\begin{tikzpicture}[baseline=(current bounding box.center), scale=0.5]
%  \node (a) at (1.5, 2) {$a$};
%  \node (b) at (3, 0) {$b$};
%  \node (c) at (0, 0) {$c$};
%  
%  \draw (a) to[bend left=50] node[midway,left]{$=$} (b); \draw (b) to[bend
%  left=50] node[midway,above]{$=$} (c); \draw (c) to[bend left=50]
%  node[midway,right]{$\neq$} (a); \end{tikzpicture}.
  
Such cycles of inconsistencies translate to cycles of inconsistent
differentials, which in turn are blocked to direct the search away from an
invalid differential path.  To do this efficiently we employ a custom-written
computer algebraic routine to detect cycles of inconsistent equations during
solving. In particular, we use a graph for finding inconsistent cycles, where in
the graph each vertex represents a variable and each edge represents a two-bit
condition. Every time a new edge is added to the graph, we search for an
inconsistent cycle involving that edge (and the shortest such cycle when one
exists).

The graph algorithm we use for detecting inconsistent cycles involves a
breadth-first search starting from vertex $v_0$ where $(v_0, v_d)$ is a newly
added edge. We look for all possible ways to reach $v_d$ excluding the edge
$(v_0, v_d)$. Each edge $(u,v)$ holds a Boolean variable $d(u,v) = u \oplus v$,
called an edge value, which tells whether the Boolean variables $u$ and $v$ are
equal or not.

For each path from $v_0$ to $v_d$ found through the method described above, we
get a cycle $v_0$, $v_1$, $\dotsc$, $v_d$, $v_0$ by adding the edge $(v_0, v_d)$
to the path. We check if there's a contradiction in a cycle (connecting Boolean
variables $v_0$ to $v_n$) by taking the $\mathbb F_2$ sum of all the edge
values, $s = d(v_0, v_1) + d(v_1, v_2) + \cdots + d(v_{d-1}, v_d)
+ d(v_d, v_0)$. If the sum $s$ is 1, there is an odd number of edges with
inequal variables, indicating an inconsistent cycle. We iterate through the
inconsistent cycles and take the shortest one for blocking.

When an inconsistency is detected it is blocked by adding a conflict clause
constructed from the parts of the SAT solver's partial assignment (during the
time of detection) implying the 2-bit conditions in the cycle.  The IPASIR-UP
interface~\cite{fazekas2023ipasir} is used for feeding the new clause to the
solver. The falsified clause causes the solver to backtrack right away, stepping
out of the invalid differential path causing the solver to backtrack earlier
than it otherwise would.

\section{Programmatic Propagation}\label{sec:prog_propagation}

During the search for a collision, we work with a partial state that comprises
known and unknown variables. Using the known variables, unknown variables may be
derived. In other words, the information that we have can spread or propagate.
This form of deduction is crucial in the search process. As mentioned in
Section~\ref{sec:encoding}, many propagation rules such as $\verb|[xx-]|
\rightarrow \verb|[-]|$ for the $\XOR$ function are encoded directly into the
SAT encoding.  However, it is not feasible to encode all possible propagation
rules on 32-bit words because there are simply too many.

There is a large body of work studying propagation in SAT encodings
and metrics by which propagation can be studied, including
propagation completeness~\cite{Bordeaux2012}, propagation strength~\cite{Brain2015},
and unit-refutation completeness~\cite{BordeauxJSM12}.  In general,
the basic unit propagation mechanism used in SAT solvers will \emph{not}
propagate all logically implied information, though IPASIR-UP supports
the inclusion of more advanced custom propagation routines.
%
%\CB{Why ``for a variant of the $\IF$ function''?}
%For instance, the difference conditions of a function's input variables can be
%used to derive the output conditions. For example, if the input conditions
%for a variant of the $\IF$ function~\eqref{eq:if} are \verb|[---]| where $x$,
%$y$, and $z$ are Boolean variables, we can propagate the output condition to a
%\verb|[-]|.}
%
%\subsection{Perfect Propagation}
%
Ideally, one would use ``perfect'' propagation encoding the most stringent
conditions possible given the current state.  For example, we describe a simple
example of perfect propagation given by \citet[Ex.~3.4]{eichlseder2013linear}.
Suppose $X=[x_3x_2x_1x_0]$ is a 4-bit word, $\Sigma(X) = (X \ggg 1) \oplus (X
\ggg 2) \oplus (X \ggg 3)=Y$, and we want to perform perfect propagation on the
differential $\wconds X \rightarrow \wconds Y$. If the differential $\wconds X$
is known to be \verb|[11--]|, then there are $2^2=4$ possibilities for $(X,X')$
because $x_2=x'_2\in\{0,1\}$ and $x_3=x'_3\in\{0,1\}$ may be chosen
independently. After trying all 4 possibilities one derives that
$(\Sigma(X),\Sigma(X'))=(Y,Y')$ must be one of $(1100,1100)$, $(0010,0010)$,
$(0001,0001)$, or $(1111,1111)$. In each case we have $Y=Y'$ meaning that we can
derive that $\wconds Y=\verb|[----]|$.

%In the previous example, the propagation of the output difference conditions can
%be done by trying all possibilities for the given differential. This technique
%of enumerating all possible values conforming to the input and output conditions
%is called perfect propagation. The set of possible values is used to construct
%the conditions in the propagated differential.
%
%In the work of \citet{eichlseder2013linear}, she described the concept of
%perfect propagation with an example (example 3.4 in her work) of a 4-bit
%$\Sigma(x) = (x \ggg 1) \oplus (x \ggg 2) \oplus (x \ggg 3)$ function. It's
%called 4-bit bitwise function because it operates on a 4-bit input and produces
%a 4-bit output.
%
%The example involves propagating (perfectly) for the differential $\wconds X
%\xrightarrow{\Sigma} \wconds Y$, where $\wconds X = \verb|[11--]|$ and $\wconds
%Y = \verb|[????]|$ are the 4-bit input and output conditions for the function.
%
%We need to enumerate all the possible input and output values conforming to
%$\wconds X$ and $\wconds X$ respectively --- involving $2^2$ operations since
%only the condition \texttt{-} has 2 possibilities (\texttt{0} and \texttt{1}).
%
%After analyzing the 4 iterations, the set of possible values within the
%constraint can be described as $\verb|[11--]| \xrightarrow{\Sigma}
%\verb|[----]|$.

Since all possibilities for ``grounding'' $\wconds X$ were explored, the maximum
amount of possible information was propagated to $\wconds Y$ and this is said to
be ``perfect'' propagation.  Unfortunately, in general perfect propagation is
infeasible because there are too many possibilities to explore.

\subsection{Bitsliced Propagation}

Perfect propagation is only feasible for small differentials with a small number
of possibilities to explore.  However, SHA-256 performs operations on 32-bit
words, which means that every function operates on 32-bit words as input. If we
want to propagate the output for a function, we'd have to deal with a relatively
large number of bits.

To keep the process computationally feasible, we only perform perfect
propagation on the output of a bitwise operation in each bit position
independently. This reduces the number of bits involved in the propagation while
still helping to deduce information. Each output condition is
propagated by enumerating all possibilities conforming to the input conditions
that the output condition is dependent on---the same as perfect propagation, but
only perfect locally. This is a practical version of perfect propagation called
``bitsliced'' propagation.

%It's also called bitsliced propagation because each step in bitwise propagation
%involves slicing off the input and output conditions from the rest of the
%conditions and finding out what it propagates to. In other words, a slice
%contains an output condition along with related input conditions, unlike perfect
%propagation that involves the all the conditions of the words. 

For example, suppose we have $X\madd Y=Z$
and we want to propagate
$\wconds X=\verb|[x-x-]|$ and $\wconds Y=\verb|[x---]|$
to $\wconds Z$.  We will focus on propagating information for
the second-least significant bit; this bitslice is highlighted in
bold in the depiction below:

%\NA{I can't make the teletype font bold.}
%\CB{There is a bold teletype font but maybe you don't have it installed.
%I'm working on Section 5.2 right now so avoiding making any changes there.}

%we can have a scenario involving propagating in modular addition.
%Our focus (in this example) is on propagating the highlighted (in bold) bitslice.
%The following is a visualization of the differential conditions in the addends
%and the sum of a modular addition example --- showing how the input conditions
%propagate to the output conditions (defined as $\wconds Y$).

\begin{center}
$
  \wconds Z =
  \begin{array}{cc}
    & \verb|?|\textcolor{red}{\textbf{\texttt{?-}}}\verb| | \\
    & \verb|[x-|\textcolor{red}{\textbf{\texttt{x}}}\verb|-]| \\
    \madd & \verb|[x-|\textcolor{red}{\textbf{\texttt{-}}}\verb|-]| \\
    \hline
    & \verb|[??|\textcolor{red}{\textbf{\texttt{?}}}\verb|-]| \\
  \end{array}
$
\end{center}

In the example above, the wordwise addition (modulo $2^4$) involves 2 addends
with the differential conditions $\verb|[x-x-]|$ and $\verb|[x---]|$, and the
first row denotes the differential conditions of the carries. In general the
bitslices involve 3 input conditions and 2 output conditions (namely, a sum
differential bit and a carry differential bit). The conditions are derived
through perfect propagation on each bitslice---in this case the slice having a
width of 1 bit.

In this example, the highlighted bitsliced differential
$\verb|[-x-??]|$ (with the last \verb|?| denoting the carry) after propagation
becomes $\verb|[-x-x?]|$ (i.e., the sum differential bit becomes an \verb|x|).
This process of bitwise propagation can be repeated for the rest of the bit
positions, resulting in propagation over a wordwise operation with a low cost.

\subsection{Wordwise Propagation}

SHA-256's hash output is calculated by a series of Boolean operations on 32-bit
words. Each step involves the state update equations of
Section~\ref{sec:stateupdate} that are used for transforming the state variables
$A$ and $E$.  We also have the message expansion equation~\eqref{eq:upd_w}
defining $W_i$ for all steps $i \ge 16$. All these equations involve modular
additions and therefore to effectively search for collisions it is essential to
have effective propagation for the modular additions. Bitsliced propagation is
helpful in deriving information for modular additions, however, this technique
doesn't capture all the relations between the bits as it is local to a bitslice
and doesn't operate on the entirety of the 32-bit words.

To mitigate this shortcoming of bitwise propagation, we utilize a global
``wordwise'' propagation technique, that is significantly
cheaper than perfect propagation on words pairs in practice but typically
derives more information than bitwise propagation.

%\subsubsection{Definition}

%Wordwise propagation is a method for propagating the differential conditions of
%individual word pairs involved in a modular addition. It
Wordwise propagation works by exploiting the
constraints in the modular addition, such as the modular integer differences of
words.  Modular differences of words were also used in the work of
\citet{wang2005break} for a different purpose.

When $A \madd B = C$ and $A' \madd B' =
C'$, wordwise propagation may derive additional information on the
differential conditions $\wconds A$ and $\wconds B$ if the modular
difference of $C$ and $C'$ is known. %$\wmdiff C \coloneqq C\msub C'$ is known.
%
%The modular difference of $A$ and $A'$ can be written as $\wmdiff A = A \msub
%A'$. Bitwise difference, on the other hand, involves bitwise operators such as
%$\XOR$ --- the bitwise difference of $A$ and $A'$ would be $A \oplus A'$.
%
Denoting modular subtraction of two 32-bit words by $\msub$, the modular
difference of $C$ and $C'$ is
\begin{equation}
  \wmdiff C \coloneqq C\msub C' = \sum_{i=0}^{31} (c_{i} - c'_i) 2^i \mod 2^{32}
  \label{eq:modular_difference}
\end{equation}
where $c_i$ and $c'_i$ denote the $i$th least significant bits of $C$ and $C'$.

In the previous example, the modular addition equations in both the hash
instances can be combined via $(A \msub A') \madd (B \msub B') = C \msub C'$
which can be rewritten as $\wmdiff A \madd \wmdiff B = \wmdiff C$.

In general, wordwise propagation is performed on equations like
\begin{equation}
  \wmdiff A_{1} \madd \wmdiff A_{2} \madd \cdots \madd \wmdiff A_n = C
  \label{eq:word_pair_mod_diffs}
\end{equation}
where $C$ can be determined in advance and we want to
derive some additional information on at least one of the
differential conditions $\wconds A_1$ to $\wconds A_n$.
%where we have $n + 1$ words $A_0$ to $A_n$. To perform wordwise
%propagation, it should be possible to derive the value of $\wmdiff A_0$ using
%equation~\eqref{eq:modular_difference} but at least one modular difference from
%$\wmdiff A_1$ to $\wmdiff A_n$ should be underivable. Wordwise propagation in
%this equation will lead to propagation of $n$ word pairs if the modular
%differences of $n$ word pairs can't be calculated.

For example, suppose we know $\delta A = \delta B$, $\wconds A=\verb|[ux-]|$,
and $\wconds B=\verb|[-n-]|$. It follows that $\delta
B=0\cdot2^2+(0-1)\cdot2+0=-2$ is known (modulo $8$), but $\delta
A=(1-0)\cdot2^2+(a_1-a'_1)\cdot2+0=4\pm2$ is either $2$ or $6$ since
$a_1-a'_1=\pm1$.  From $\wconds A$ alone the value of $\delta A$ cannot be
determined exactly, but when the additional constraint $\delta A=\delta B$ is
considered it is clear that the only solution is $\delta A=6$ meaning that
$a_1-a'_1=1$.  Thus, wordwise propagation in this case would derive $\wconds
A=\verb|[u|\textcolor{red}{\textbf{\texttt{u}}}\verb|-]|$.

To avoid dealing with negative numbers, the differential conditions \verb|x|,
\verb|n|, and \verb|?| are normalized by adding an appropriate power of two. For
example, in the above example $2^1$ would be added making the equation
\[ (1-0)\cdot2^2 + w\cdot2^1 + 0 = -2 + 2^1 = 0 \qquad\text{where $w\coloneqq
a_1-a'_1+1\in\{0,2\}$} \] becoming $(1+v)\cdot 2^2=0$ where $v\coloneqq
w/2\in\{0,1\}$.  As a $3$-bit bitvector equation (hence performed modulo $2^3$),
this is $[1+v,0,0]=[0,0,0]$ which has just one solution $v=1$.

\paragraph{Dividing into subproblems.}

After reducing equations of the form~\eqref{eq:word_pair_mod_diffs} to bitvector
equations, we want to determine all solutions for the variables.  To do so, we
search for possible values through brute force.  Since this has an exponential
time complexity, we divide the problem into smaller components by analyzing the
cascading effects of the carries.  In practice, this reduces the computational
cost significantly as the subproblems can usually be solved quickly, and any
subproblem that is too expensive to solve can be skipped without affecting the
other subproblems. In our work, we consider any subproblem with more than 10
variables as expensive, limiting the number of (brute force) iterations to
$2^{10} = 1024$ for a subproblem.

%\CB{If variables' values can be fixed then
%they don't contribute to increasing the search space of the subproblem.}
%\NA{Did you mean fixing the values without brute force?}
%\CB{I thought maybe there were cases where variables in a subproblem
%that did not contribute to increasing the search space, but maybe that's
%not the case.}

As an example, suppose $\wmdiff A \madd \wmdiff B = \wmdiff C$ where
$\wconds A = \verb|[u1xxx]|$, $\wconds B = \verb|[xx-nx]|$,
and $\wconds C = \verb|[---u-]|$.
Then $\delta C=2$, and after normalizing $A$ and $B$ we derive
\[ (\delta A+2^2+2+1) \madd (\delta B + 2^4+2^3+2+1) = \delta A \madd \delta B \madd 2 = 4 , \]
becoming the bitvector modular addition problem
\begin{center}
  $
  \wmdiff A \madd \wmdiff B \madd 2 =
  \begin{array}{@{}c@{\hspace{0.1cm}}c@{\hspace{0.1cm}}c@{\hspace{0.1cm}}c@{\hspace{0.1cm}}c@{\hspace{0.1cm}}c@{\hspace{0.1cm}}c@{\hspace{0.1cm}}c@{}}
    & [ & 1, & v_3, & v_2, & v_0, & 0 & ] \\
    \madd & [ & v_4, & 0, & 0, & v_1, & 0 & ] \\
    \hline
    & & 0 & 0 & 1 & 0 & 0 &
  \end{array}
  $
\end{center}
where $v_0$, $\dotsc$, $v_4\in\{0,1\}$.

Now, a linear scan is performed starting from the least significant digit (the
rightmost column). Since there's no variable in the first column, we skip it.
Next, we have a subproblem candidate $v_0 + v_1 \equiv 0\pmod{2}$.  Since this
addition can overflow, the next column must be included in the subproblem.
Including the next column results in the subproblem
$v_0+v_1+2v_2\equiv2\pmod{4}$ which cannot overflow since $v_0=v_1=v_2=1$ is not
a solution. Thus, the problem starting in the next column is independent.

The next problem is then $v_3\equiv0\pmod{2}$, which only has one solution
$v_3=0$ and thus also cannot overflow and is independent of the final column.
Similarly, the final column results in the subproblem $1+v_4\equiv0\pmod{2}$
which only has the single solution $v_4=1$.

In this example wordwise propagation thus determines that $v_4=1$ and $v_3=0$.
Since $v_4$ arose from the second `\verb|x|' in $\wconds B$ which encodes the
difference $b_3\oplus b'_3$,
%\NA{Do we not have to define $b_3$, $b'_3$, and similar variables?} \CB{I think
%they are fairly implicit, but we can be a bit more explicit.}
i.e., $v_4=(b_3-b'_3+1)/2$, we derive that $(b_3,b'_3)=(1,0)$. Similarly, $v_3$
arose from the `\verb|x|' in $\wconds A$ encoding $a_2\oplus a'_2$, i.e.,
$v_3=(a_2-a'_2+1)/2$, and we derive that $(a_2,a'_2)=(0,1)$.  Thus, in this
example wordwise propagation deduces the updated differential conditions
$\wconds A=\verb|[u1|\textcolor{red}{\textbf{\texttt{n}}}\verb|xx]|$ and
$\wconds B=\verb|[x|\textcolor{red}{\textbf{\texttt{u}}}\verb|-nx]|$.

\paragraph{Implementation Details.}

Wordwise propagation was applied to all the modular addition equations,
specifically the message expansion~\eqref{eq:upd_w} and state update transformation equations,
including the one for the auxiliary word $T_i$.  However, the only
variables that were propagated were the differential variables
in $\wconds A_i$, $\wconds E_i$, and $\wconds W_i$.

During the wordwise propagation routine, a heuristic that we used
which we found dramatically improved the efficiency of the solver
was to assume that any differential `\texttt{?}' in the auxiliary
variables (including $\wconds T_i$ and the differential variables corresponding
to the output of $\IF$, $\MAJ$, $\sigma_0$, $\Sigma_0$, etc.)\ was
actually a `\verb|-|' differential.
%
%In practice, assuming the differential condition `\texttt{?}' to be `\texttt{-}'
%for the auxiliary word pairs (anything other than $\wconds A$, $\wconds E$, and
%$\wconds W$) clearly increased the number of propagations and significantly
%reduced the time needed to find SFS collisions. Without these assumptions,
%wordwise propagation didn't seem to help the solver for any number of steps in
%our benchmark, which ranged from 20 to 38.
%
In practice, making this assumption allowed the modular difference of the
auxiliary differential words to be calculable much more frequently,
and increased the likelihood that variables in the
word differentials $\wconds A_i$, $\wconds E_i$, and $\wconds W_i$ were derived.
This heuristic is related to the `decision' search strategy of \citet{mendel2011finding}
which always first imposes a `\verb|-|' for a `\verb|?|' before imposing a `\verb|x|'.
However, we found that making
assumptions on the primary word differentials $\wconds A_i$, $\wconds E_i$, and
$\wconds W_i$ themselves significantly decreased the solver's performance, preventing us
from finding SFS collisions for SHA-256 beyond 28 steps.

\section{Implementation and Results}\label{sec:results}

Our programmatic SAT solver was implemented in CaDiCaL
1.8.0~\cite{biere2023cadical} using the programmatic interface
IPASIR-UP~\cite{fazekas2023ipasir}. Our experiments were run in the Digital
Research Alliance of Canada's~\cite{baldwin2012compute} Narval cluster.  Each
SAT solver instance ran on a single core of an AMD Rome 7532 processor running
at 2.4~GHz %(Graham) or Intel Gold 6148 Skylake @ 2.4 GHz
with 4~GiB of RAM\@. %---we found very similar CPU times for the same set
%of single-threaded instances on both CPU models.
Our implementation is free software and is available
online.\footnote{\url{https://github.com/nahiyan/cadical-sha256}}

\subsection{Implementation}

%Our framework comprises CaDiCaL 1.8 that includes the programmatic interface,
%IPASIR-UP.
The IPASIR-UP programmatic interface provides access to the current state of the
solver for the relevant variables (i.e., those encoding the state of the hash
function and the differential variables). IPASIR-UP also enables us to perform
custom propagation and branching as well as learning custom conflict clauses.
%a trail of the solver for the
%variables of our focus, which we use for constructing a state of the variables
%defined through the differential conditions. We perform custom propagation,
%branching, and learning using the interface based on our high-level state.

For implementing inconsistency blocking, we used our own implementation of a
graph library.  This houses the graph algorithm described in
Section~\ref{sec:prog_incons_blocking} for detecting minimal inconsistent
cycles in linear time.
%For implementing wordwise propagation,
Our implementations of bitsliced and wordwise propagation, as well as the
two-bit condition detection engine used for inconsistency blocking, are based on
the ideas described in the works of \citet{mendel2011finding} and
\citet{eichlseder2013linear,eichlseder2018differential}. We used a
least-recently-used cache data structure for caching propagation rules and rules
for deriving two-bit conditions. The cache doesn't grow beyond the maximum
available RAM, since the least frequently used entries are deleted on the fly.

In our experiments, most queries to the propagation and two-bit detection
engines could be served from the cache, which is much faster than deriving the
propagation rules or the two-bit conditions on the fly each time. Since the
set of rules that are queried throughout the entire runtime is usually small
(i.e., consumes a small portion of the total CPU time), it wasn't necessary to
precompute any rules.

%\paragraph{Wordwise programmatic propagation.}
%
%\CB{This belongs with the description of the implementation, not the method.}

We perform bitsliced propagation for all operations in SHA-256 (including the
modular additions) alongside the solver's built-in Boolean constraint
propagation (BCP). As wordwise propagation is much more expensive in terms of
computational cost, it's performed only when the SAT solver finishes with the
other propagation methods. This way, wordwise propagation only deduces the
conditions that bitsliced propagation couldn't.

%We use IPASIR-UP for external propagation. The interface lets us write callback
%functions during the different stages of solving --- propagation, branching, and
%learning.
IPASIR-UP asks for a ``reason'' clause of propagated literals when it becomes
necessary for the solver to know why a literal was propagated. For bitsliced
propagation, these reason clauses were relatively short, so in this case we
provided reason clauses directly via IPASIR-UP's interface.  On the other hand,
wordwise propagation involves multiple word pairs and a single propagated
literal may depend on a relatively large number of bits, leading to long reason
clauses.  To avoid overwhelming the solver with long reason clauses, we did not
use IPASIR-UP's propagation interface for wordwise propagation and instead set
the values of any literals deduced by wordwise propagation via branching.

\subsection{Results}

We performed the same experiment with three separate solvers: an unmodified
version of CaDiCaL 1.8.0, a version of CaDiCaL with programmatic bitsliced and
wordwise propagation, and a version of CaDiCaL with both programmatic
propagation and inconsistency blocking. We also tried using CryptoMiniSat
5.11.21~\cite{Soos2009}, given that it supports XOR constraints natively and has
been tuned to work on cryptographic problems.  However, we did not pursue this
extensively as CryptoMiniSat currently does not have a programmatic interface
and did not perform as well as CaDiCaL.

\begin{figure}
  \centering
  \includegraphics[scale=0.75]{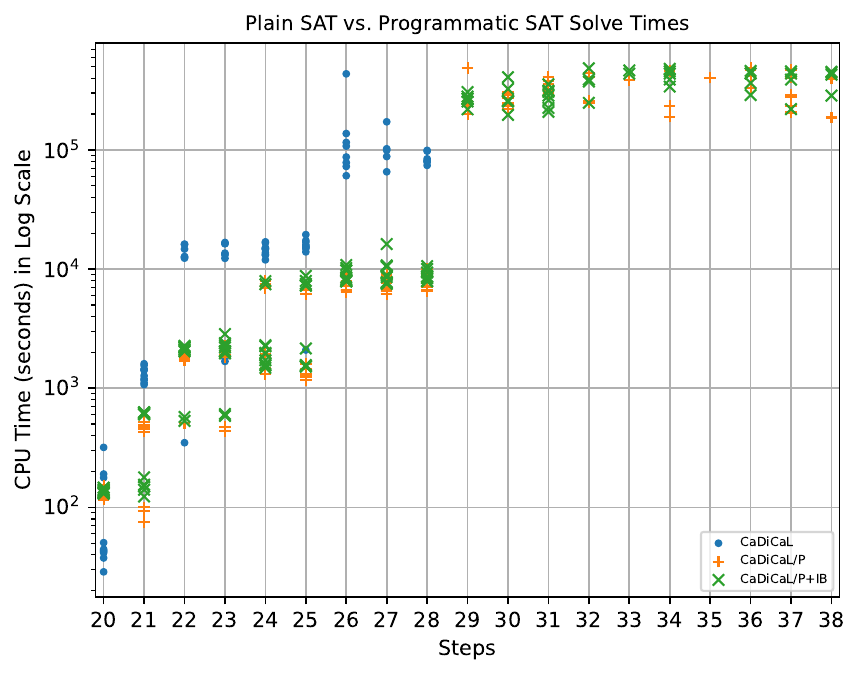}
  \caption{Running times for finding a SFS collision for step-reduced SHA-256
  for a varying number of steps.
  The plot compares a plain SAT solver with two programmatic SAT solvers.
  The lack of a data point indicates no collisions were found
  within 500,000 seconds.
  %The programmatic approach clearly appears to outperform plain SAT both in terms of
  %the runtime and the maximum number of steps reached.
  %Plain SAT could reach up
  %to 28 steps while programmatic SAT went up to 38.
  }\label{fig:sat-v-psat-20-38}
\end{figure}

With each solver we searched for semi-free-start collisions for step-reduced
SHA-256 with 20 to 38 steps. In order to reduce the randomness inherent in the
search, each instance was solved ten times independently using 10 different
random seeds, though these 10 different seeds were consistently used across all
our experiments. Each instance was run for a time limit of 500,000 seconds
(roughly 5.8 days). The number of instances successfully solved in each case is
given in Table~\ref{tbl:collisions_count}, and the minimum time for finding a
collision is plotted in Figure~\ref{fig:sat-v-psat-20-38}. The starting points
used for the %instances with
%21 steps, 25 steps, 28 steps, and 38 steps are given
21-step, 25-step, 28-step, and 38-step instances are given in the appendix.  The
starting points for all other step counts were formed from one of these starting
points by dropping a number of rows at the bottom, e.g., the 26-step instance
matches the 28-step starting point with two rows removed. This means that the
instances in the step ranges 20--21, 22--25, 26--28, and 29--38 can be expected
to roughly have similar difficulty as they were created from the same starting
point.

The results show that programmatic propagation was clearly
effective at helping the solver find SFS collisions.
The plain SAT solver could only find SFS collisions up to 28 steps,
% while CaDiCaL with programmatic propagation found SFS collisions up to 37 steps,
% and CaDiCaL with both programmatic propagation and inconsistency blocking was
% able to find SFS collisions for every number of steps from 20 up to and
% including 38 steps.
while CaDiCaL with programmatic propagation, both with and without inconsistency
blocking, successfully found SFS collisions for every step count from 20 to 38,
with the exception of 35---no 35-step instances were solved when both
programmatic propagation and inconsistency blocking were enabled (see
Table~\ref{tbl:collisions_count}).
In general, we found inconsistency blocking tended to decrease the efficiency of the solver,
although using inconsistency blocking did result in the fastest
solve times for the instances with 30, 31, 32, and 36 steps.

%Additionally, only for the instances with 36 steps did enabling inconsistency blocking
%cause the number of SFS collisions found to increase (see
%Table~\ref{tbl:collisions_count}).
%Moreover, the number of SFS collisions found (see
%Table~\ref{tbl:collisions_count}) didn't increase in most cases with the
%combination of programmatic propagation and inconsistency blocking, compared to
%only programmatic propagation.

% \CB{It doesn't seem like inconsistency blocking was effective. Some kind of
% comparison should be done.}

%\NA{I think we can replace ``programmatic SAT+CAS'' with ``programmatic SAT''.}

\begin{table}
  \centering
  \caption{Number of step-reduced SFS collisions found in each instance for the
  3 methods: plain CaDiCaL, CaDiCaL with Propagation (P), and CaDiCaL with
  Propagation and Inconsistency Blocking (P+IB). For each number of steps and
  solver, we solved the same instance using 10 different SAT solver seeds.}
  \label{tbl:collisions_count}
  \setlength{\tabcolsep}{3.8pt}
  \begin{tabular}{*{20}{r}}
    \toprule
    \textbf{Steps} & 20 & 21 & 22 & 23 & 24 & 25 & 26 & 27 & 28 & 29 & 30 & 31 &
    32 & 33 & 34 & 35 & 36 & 37 & 38 \\
    \midrule
    \textbf{CaDiCaL} & 10 & 10 & 10 & 10 & 10 & 10 & 8 & 5 & 7 & 0 & 0 & 0 & 0 &
    0 & 0 & 0 & 0 & 0 & 0 \\
    \textbf{CaDiCaL/P} & 10 & 10 & 10 & 10 & 10 & 10 & 10 & 10 & 10 & 6 & 9 & 8
    & 4 & 2 & 3 & 1 & 4 & 7 & 3 \\
    \textbf{CaDiCaL/P+IB} & 10 & 10 & 10 & 10 & 10 & 10 & 10 & 10 & 10 & 6 & 7 &
    7 & 4 & 2 & 5 & 0 & 4 & 4 & 4 \\
    \bottomrule
  \end{tabular}
\end{table}

%The SAT solver that we've used, CaDiCaL 1.8.0, could find SFS collisions up to
%28 steps. However, the programmatic techniques not only sped up the performance
%of the solver but reached up to 38 steps within a time limit of
%\numprint{300000} seconds (roughly 3.5 days). The 38-step SFS
%collision is provided in the appendix. The experiments were run
%in Digital Research Alliance of Canada's~\cite{baldwin2012compute} Graham
%cluster --- each SAT solver instance ran on a single CPU core of Intel E5\-2683
%v4 Broadwell @ 2.1GHz with 4 GB of RAM.

%More specifically, programmatic propagation (bitwise + wordwise) could derive
%collisions up to 37 steps. Combining the programmatic propagation with
%inconsistency blocking added some overhead but took the attack further to 38
%steps. Figure~\ref{fig:sat-v-psat-20-38} shows the best times for each
%instance.
%
%For each solver and step count, we performed the benchmark on 10 different
%seeds. This led to the solver initiating from different places in the search
%space and therefore take a distinct path to the solution. Having multiple seeds
%for the same solver and step count also enabled us to (in some sense)
%parallelize the process and utilize multiple CPU threads (the solver is
%single-threaded). Moreover, multiple seeds in most cases gave unique SFS
%collisions, which helped us get an idea of the difficulty of a specific
%configuration. Table~\ref{tbl:collisions_count} shows the number of collisions
%found for each method and order.

\section{Conclusion}\label{sec:conclusion}

In this work we combine the programmatic SAT+CAS paradigm with the differential
cryptanalysis techniques used in previous collision attacks on SHA-256.  In the
process, we demonstrate that these computer algebraic techniques can
dramatically improve the performance of the SAT solver, enabling the SAT+CAS
solver to find a semi-free-start collision of SHA-256 with 38 steps, while a
plain SAT solver could go no further than 28 steps.  Moreover, previous 38-step
SFS collisions~\cite{mendel2013improving} were found with a highly sophisticated
search tool specifically written to find SHA-256 collisions, while our work used
the general purpose SAT solver CaDiCaL coupled with the IPASIR-UP
interface~\cite{fazekas2023ipasir} for custom propagation, branching, and
learning.  Thus, we were able to exploit the power of modern SAT solvers without
needing to write a search tool from scratch.

At the time the work in this paper was performed, the best SFS collision ever
found for SHA-256 contained 38 steps~\cite{mendel2013improving}. Just prior to
submitting this work, the authors became aware of the work of \citet{Li2024}
appearing at EUROCRYPT~2024 that finds for the first time a 39-step SHA-256
SFS collision.  \citet{Li2024} also use a SAT-based approach, but with a
significantly different encoding.  Determining if the
SAT+CAS approach can also be useful with this alternate encoding will be the
subject of future work.

%previous SAT-based
%
%outperformed a plain SAT solving approach. Leveraging CaDiCaL's IPASIR-UP
%interface, we have reached up to 38 steps, while the plain SAT approach could go
%no further than 28 steps. This sets a new benchmark in utilizing a SAT solver
%for finding SHA-256 SFS collisions. An advantage of our method is that existing
%highly-tuned SAT solvers can be used instead of writing a dedicated search tool
%to find collisions.
%
%However, we couldn't extend any SFS collision to a full collision beyond 25
%steps, which remains for the future. While there are potential rooms for
%improvements in our programmatic techniques and the extension technique, it lays
%the foundation of differential cryptanalysis with programmatic SAT.

\section*{Acknowledgements}

We thank the chairs of the SC-Square 2024 workshop, Daniela Kaufmann and Chris Brown,
for their flexibility during the publication process of this paper, and the anonymous reviewers
for their detailed comments.
We also thank Maria Eichlseder for her insight and answering a number of questions
concerning state-of-the-art hash function collision search tools, as well as
Oleg Zaikin for answering questions about his work on inverting 43-step
MD4~\cite{Zaikin2022}.

\bibliography{refs}

\pagebreak
\appendix
\section*{Appendix}

% In the appendix we provide an example of a 38-step semi-free-start SHA-256
% collision that we found (Table~\ref{tbl:sfs_coll_38}) along with the explicit
% starting points that we used in our search.
In the appendix, we provide an example of a 38-step semi-free-start SHA-256
collision that we found (Table~\ref{tbl:sfs_coll_38}), a table showing
its differential characteristic (Table~\ref{tbl:dc_38}), and the starting points
that we used in our search.

The 21-step starting point (Table~\ref{tbl:sp_21}) is taken from the work of
\citet{prokop2016differential}. The starting point for 25 steps
(Table~\ref{tbl:sp_25}) is an extended version of the 24-step starting point
provided by \citet{prokop2016differential}. The 28-step starting point
(Table~\ref{tbl:sp_28}) is a slightly modified version of the starting point
used by \citet{mendel2013improving}.

The 38-step starting point (Table~\ref{tbl:sp_38}) is constructed based on the
38-step differential characteristic provided by \citet{mendel2013improving}---in
particular the differential words $\wconds W_{15}$, $\wconds W_{23}$, $\wconds
W_{24}$, $\wconds A_{15}$, and $\wconds A_{16}$.  The `\verb|x|'s in these words
are placed under the heuristic assumption that these words have a low (but
nonzero) Hamming weight. The differential word $\wconds W_{24}$ (with a Hamming
weight of 1 and an `\verb|x|' in position 2) was taken from their starting
point.  This propagates to the `\verb|x|'s in $\wconds W_{15}$ and $\wconds
A_{16}$ (and both those words were assumed to have a Hamming weight of 1 as
well) as well as the `\verb|x|' in position 16 of $\wconds W_{23}$. Then setting
position 16 of $\wconds A_{15}$ to `\verb|x|' causes it to propagate to $\wconds
E_{19, 16}$ and cancel out with $\wconds W_{23, 16}$ in the state update
transformation equation of $T_{23}$ (and similarly for position~27 of $\wconds
A_{15}$). $\wconds W_{23,27}$ is set to `\texttt{x}' to cancel out with
$\sigma_0(W_{15})$ in step 30 of the message expansion equation.

\begin{table}[h]
  \caption{SFS collision for 38 steps found with programmatic propagation and
    inconsistency blocking. $h_0$ is the chaining value, $(M, M')$ is the
    colliding message pair, and $h_1$ is the hash of $M$ and $M'$. Word pairs in
    $M$ and $M'$ that have differences are enclosed in a box.}
  \label{tbl:sfs_coll_38}
  \begin{tabular}{|c|p{13.5cm}|}
    \hline
    \textbf{$h_0$} & \texttt{afea2566 1e0a73e2 da747de7 34381a7f 6f4c0d98
    897dd98c 592ba6ad 2aa5e80} \\
    \hline
    \textbf{$M$} & \texttt{5b5058d2 901f87fb 254bcfa2 5f8d7dc1 fb1053be 0622e1f8
    da8801c2 \setlength{\fboxsep}{0.02cm}\fbox{\textcolor{red}{a951cfb}b} \fbox{5db4\textcolor{red}{2ffd}}
    683b4391 \fbox{f8\textcolor{red}{7ea}bbd} e928b976 3675cc55 6ebe78be e3031536
    \fbox{c2de906\textcolor{red}{f}}} \\
    \hline
    \textbf{$M'$} & \texttt{5b5058d2 901f87fb 254bcfa2 5f8d7dc1 fb1053be 0622e1f8
    da8801c2 \setlength{\fboxsep}{0.02cm}\fbox{\textcolor{red}{9737d17}b} \fbox{5db4\textcolor{red}{3001}}
    683b4391 \fbox{f8\textcolor{red}{812}bbd} e928b976 3675cc55 6ebe78be e3031536
    \fbox{c2de906\textcolor{red}{b}}} \\
    \hline
    \textbf{$h_1$} & \texttt{d0e019f7 408269d3 24296a7b 30df8e7f 95d2bff8
    34e2bca6 6c50a294 ddb4254a} \\
    \hline
  \end{tabular}
\end{table}

\begin{table}
%<<<<<<< HEAD
%  \caption{Differential characteristics for the 38-step SFS collision presented
%  in Table~\ref{tbl:sfs_coll_38}. All word pairs with differences (having
%  \texttt{u} or \texttt{n} conditions) are enclosed in a box, with the specific
%  \texttt{u} and \texttt{n} conditions highlighted in red.}
%=======
  \caption{The differential characteristic for the 38-step semi-free-start collision presented
  in Table~\ref{tbl:sfs_coll_38}.  The words with a nonzero difference (i.e., including a
  `\texttt{u}' or `\texttt{n}' differential) are enclosed in a box.
  Interestingly, compared to the 38-step semi-free-start collision presented by
  \citet{mendel2013improving},
  an additional two words ($\wconds A_8$ and $\wconds E_{10}$) have a zero difference.}
%>>>>>>> 9ebcbe5 (Updates to table 5 (differential characteristic of 38-step SFS collision))
  \label{tbl:dc_38}
  \scalebox{0.7}{
    \begin{tabular}{|r|c|c|c|}
      \hline
      $i$ & $\wconds A_i$ & $\wconds E_i$ & $\wconds W_i$ \\
      \hline
      \DCXXXVIII
      \hline
    \end{tabular}
  }
\end{table}

\begin{table}
  \caption{Starting point for a 21-step semi-free-start collision.}
  \label{tbl:sp_21}
  \scalebox{0.7}{
    \begin{tabular}{|r|c|c|c|}
      \hline
      $i$ & $\wconds A_i$ & $\wconds E_i$ & $\wconds W_i$ \\
      \hline
      \SPXXI
      \hline
    \end{tabular}
  }
\end{table}

\begin{table}
  \caption{Starting point for a 25-step semi-free-start collision.}
  \label{tbl:sp_25}
  \scalebox{0.7}{
    \begin{tabular}{|r|c|c|c|}
      \hline
      $i$ & $\wconds A_i$ & $\wconds E_i$ & $\wconds W_i$ \\
      \hline
      \SPXXV
      \hline
    \end{tabular}
  }
\end{table}

\begin{table}
  \caption{Starting point for a 28-step semi-free-start collision.}
  \label{tbl:sp_28}
  \scalebox{0.7}{
    \begin{tabular}{|r|c|c|c|}
      \hline
      $i$ & $\wconds A_i$ & $\wconds E_i$ & $\wconds W_i$ \\
      \hline
      \SPXXVIII
      \hline
    \end{tabular}
  }
\end{table}

\begin{table}
  \caption{Starting point for a 38-step semi-free-start collision.}
  \label{tbl:sp_38}
  \scalebox{0.7}{
    \begin{tabular}{|r|c|c|c|}
      \hline
      $i$ & $\wconds A_i$ & $\wconds E_i$ & $\wconds W_i$ \\
      \hline
      \SPXXXVIII
      \hline
    \end{tabular}
  }
\end{table}

\end{document}